\documentclass[twocolumn]{aastex631}
\usepackage{xcolor}
\usepackage{amsmath}

\def\sn{SN\,2023ixf}

\def\xmm{\emph{XMM-Newton}}

\def\chandra{\emph{CXO}}
\def\nustar{\emph{NuSTAR}}

\begin{document}

%\title{Multi-wavelength emission from supernova 2023ixf (Paper I)}
\title{Dinosaur in a Haystack \footnote{Title inspired by a news report on the re-discovery of fossilized dinosaur eggs in rural Gujarat in Indian Express  28 October 2024.}: X-ray View of the Entrails of SN 2023ixf and the Radio Afterglow of Its Interaction with the Medium Spawned by the Progenitor Star (Paper 1)}

\author[0000-0002-8070-5400]{Nayana A. J.}
\affiliation{Department of Astronomy, University of California, Berkeley, CA 94720-3411, USA}

\author[0000-0003-4768-7586]{Raffaella Margutti}
\affiliation{Department of Astronomy, University of California, Berkeley, CA 94720-3411, USA}
\affiliation{Department of Physics, University of California, 366 Physics North MC 7300,
Berkeley, CA 94720, USA}

\author[0000-0000-0000-0000]{Eli Wiston}
\affiliation{Department of Astronomy, University of California, Berkeley, CA 94720-3411, USA}

\author[0000-0002-7706-5668]{Ryan Chornock} 
\affiliation{Department of Astronomy, University of California, Berkeley, CA 94720-3411, USA}

\author[0000-0001-6278-1576]{Sergio Campana}
\affiliation{INAF--Osservatorio Astronomico di Brera, Via Bianchi 46, I-23807, Merate (LC), Itay}

\author[0000-0003-1792-2338]{Tanmoy Laskar}
\affiliation{Department of Physics \& Astronomy, University of Utah, Salt Lake City, UT 84112, USA}

\author[0000-0002-5358-5642]{Kohta Murase}
\affiliation{Department of Physics, Department of Astronomy \& Astrophysics, \& Center for Multimessenger Astrophysics, Institute for Gravitation \& the Cosmos, The Pennsylvania State University, University Park, PA 16802, USA}
\affiliation{Yukawa Institute for Theoretical Physics, Kitashirakawa-Oiwake-cho, Sakyo-ku, Kyoto, 606-8502, Japan}

\author[0000-0001-6971-4851]{Melanie Krips}
\affiliation{Institut de Radioastronomie Millimétrique (IRAM), 300 Rue de la Piscine, 38400 Saint-Martin-d'H$\grave{e}$res, France}

\author[0000-0002-0786-7307]{Giulia Migliori}
\affiliation{INAF Istituto di Radioastronomia, via Gobetti 101, 40129 Bologna, Italy.}

\author[0000-0002-6347-3089]{Daichi Tsuna}
\affiliation{TAPIR, Mailcode 350-17, California Institute of Technology, Pasadena, CA 91125, USA}
\affiliation{Research Center for the Early Universe (RESCEU), School of Science, The University of Tokyo, 7-3-1 Hongo, Bunkyo-ku, Tokyo 113-0033, Japan}

\author[0000-0002-8297-2473]{Kate~D.~Alexander}
\affiliation{Department of Astronomy/Steward Observatory, 933 North Cherry Avenue, Rm. N204, Tucson, AZ 85721-0065, USA}

\author[0000-0002-0786-7307]{Poonam Chandra}
\affiliation{National Radio Astronomy Observatory,
520 Edgemont Rd, Charlottesville VA 22903, USA}

\author[0000-0002-0592-4152]{Michael Bietenholz}
\affiliation{SARAO/Hartebeesthoek Radio Observatory, PO Box 443, Krugersdorp 1740, South Africa}

\author[0000-0002-9392-9681]{Edo Berger}
\affiliation{Center for Astrophysics \textbar{} Harvard \& Smithsonian, 60 Garden Street, Cambridge, MA 02138-1516, USA}

\author[0000-0002-9117-7244]{Roger A. Chevalier}
\affiliation{Department of Astronomy, University of Virginia,
 Charlottesville VA 22904-4325, USA}

\author[0000-0002-3137-4633]{Fabio De Colle}
\affiliation{Instituto de Ciencias Nucleares, Universidad Nacional Aut\'{o}noma de M\'{e}xico, Apartado Postal 70-264, 04510 M\'{e}xico, CDMX, Mexico} 

 \author[0000-0003-0599-8407]{Luc Dessart}
\affiliation{Institut d'Astrophysique de Paris, CNRS-Sorbonne Universit\'e, 98 bis boulevard Arago, 75014 Paris, France}

\author[0000-0002-6679-0012]{Rebecca Diesing}
\affiliation{School of Natural Sciences, Institute for Advanced Study, Princeton, NJ 08540, USA}

\author[0000-0002-1984-2932]{Brian W. Grefenstette}
\affiliation{Space Radiation Laboratory
California Institute of Technology 
1200 E California Blvd 
Pasadena, CA 91125, USA}

\author[0000-0002-3934-2644]{Wynn V. Jacobson-Gal\'{a}n}
\altaffiliation{NSF Graduate Student Fellow}
\affiliation{Department of Astronomy, University of California, Berkeley, CA 94720-3411, USA}

\author[0000-0003-2611-7269]{Keiichi Maeda}]
\affiliation{Department of Astronomy, Kyoto University, Kitashirakawa-Oiwake-cho, Sakyo-ku, Kyoto, 606-8502, Japan}

\author[0000-0001-9814-2354]{Benito Marcote}
\affiliation{Joint Institute for VLBI ERIC, Oude Hoogeveensedijk 4, 7991~PD Dwingeloo, The Netherlands}
\affiliation{ASTRON, Netherlands Institute for Radio Astronomy, Oude Hoogeveensedijk 4, 7991~PD Dwingeloo, The Netherlands.}

\author[0000-0002-4513-3849]{David Matthews}
\affiliation{Illinois Institute of Technology Department of Physics, Chicago IL 60616 USA}

\author[0000-0002-0763-3885]{Dan Milisavljevic}
\affiliation{Purdue University, Department of Physics and Astronomy, 525 Northwestern Ave, West Lafayette, IN 47907 }
\affiliation{Integrative Data Science Initiative, Purdue University, West Lafayette, IN 47907, USA}

\author[0000-0002-3356-5855]{Alak K. Ray}
\affiliation{Homi Bhabha Centre for Science Education, TIFR, Mumbai 400088, India }

\author[0000-0003-4254-2724]{Andrea Reguitti}
\affiliation{INAF – Osservatorio Astronomico di Brera, Via E. Bianchi 46, I-23807 Merate (LC), Italy
}
\affiliation{INAF – Osservatorio Astronomico di Padova, Vicolo dell'Osservatorio 5, I-35122 Padova, Italy}

\author[0000-0002-5283-933X]{Ava Polzin}
\affiliation{Department of Astronomy  \& Astrophysics, The University of Chicago, Chicago, IL 60637 USA}
\begin{abstract}
We present the results from our extensive hard-to-soft X-ray (NuSTAR, Swift-XRT, XMM-Newton, Chandra) and meter-to-mm wave radio (GMRT, VLA, NOEMA) monitoring campaign of the very nearby (d$=6.9$\,Mpc) Type II \sn{} spanning $\approx$\,4--165\,d post-explosion. This unprecedented dataset enables inferences on the explosion's circumstellar medium (CSM) density and geometry. Specifically, we find that the luminous X-ray emission is well modeled by thermal free-free radiation from the forward shock with rapidly decreasing photo-electric absorption with time. The radio spectrum is dominated by synchrotron radiation from the same shock, and the NOEMA detection of high-frequency radio emission may indicate a new component consistent with the secondary origin. Similar to the X-rays, the level of free-free absorption affecting the radio spectrum rapidly decreases with time as a consequence of the shock propagation into the dense CSM. While the X-ray and the radio modeling independently support the presence of a dense medium corresponding to an \emph{effective} mass-loss rate $\dot{M} \approx 10^{-4}\, \rm M_{\odot}\,yr^{-1}$ at $R = (0.4-14) \times 10^{15}$ cm (for $v_{\rm w}=\rm 25 \,km\,s^{-1}$), our study points at a complex CSM density structure with asymmetries and clumps. The inferred densities are $\approx$10--100 times those of typical red supergiants, indicating an extreme mass-loss phase of the progenitor in the $\approx$200 years preceding core collapse, which leads to the most X-ray luminous Type II SN and the one with the most delayed emergence of radio emission. These results add to the picture of the complex mass-loss history of massive stars on the verge of collapse and demonstrate the need for panchromatic campaigns to fully map their intricate environments.
\end{abstract}


%%%%%%%%%%%%%%%%%%%%%%%%%%%%%%%%%%%%%%%%%%%%%%%%%%%%%%%%%
\section{Introduction} \label{sec:intro}
Observations of core-collapse supernovae (CCSNe) across optical, radio, and X-ray wavelengths have provided strong evidence for a time-dependent mass-loss history of massive stars in the centuries preceding core-collapse \citep{Smith14,Chevalier17}. The consequent circumstellar medium (CSM) would be complex, with significant structure in the immediate environments and at farther distances from the explosion site. While the time-dependent mass-loss of massive stars is not well understood by the current stellar evolution models, understanding the structure of the CSM could put valuable constraints on various proposed mechanisms of mass-loss (e.g., \citealt{Quataert2012,Quataert2016,Owocki2017,Fuller2017}). 

Various observational techniques that probe the CSM at different distances from the explosion center include direct detection of pre-explosion stellar outbursts (e.g., \citealt{Kilpatrick2018,Jacobson-Galan2022,Reguitti2024,Strotjohann2021}); the presence of narrow emission lines in early optical spectra known as ``flash spectroscopy'' features (e.g., \citealt{Khazov2016,Yaron2017,Boian2020,Terreran2022,Bruch2021,Bruch2023,JacobsonGalan2024a-sample-paper,JacobsonGalan2024b-sn2024ggi,Leonard2000}); or from the properties of early optical light curves (e.g., \citealt{Das2017,Forster2018}). Outside the UV/optical spectrum, the shock interaction with a dense medium manifests as luminous X-ray and radio emission (e.g., \citealt{Chevalier17} for a recent review), where X-ray emission typically arises from thermal bremsstrahlung and radio emission is due to shock-accelerated relativistic electrons producing synchrotron radiation. By analyzing the spectral and temporal evolution of X-ray and radio emission, the CSM density profile can be inferred (e.g., \citealt{Chakraborti13ej,Chakraborti14dj,Chandra12,Dwarkadas14}  for type-II SNe).

\sn{} was discovered by K. Itagaki on 2023 May 19, in the galaxy M101 (NGC 5457). The SN was classified as a type II based on the optical spectrum \citep{perley2023-tns}. We adopt a distance of 6.9 Mpc \citep{Riess22} and a time of first light corresponding to MJD 60082.833 \citep{Mao23}. All times are reported with respect to the time of first light. The small uncertainty ($\approx 0.02$ days) on the time of first light has no impact on our major conclusions. 

Extensive data has been collected on \sn{} and its explosion site across the electromagnetic spectrum, including early optical/UV observations sampling the light curve rise \citep{Zimmerman23,Hosseinzadeh23}, early to late-time optical spectroscopic observations revealing flash ionization features at early times \citep{JacobsonGalan23,smith2023,Teja23,Yamanaka23,Hiramatsu23,Bostroem23,Avinash2024}, pre-explosion image analysis in optical and IR bands \citep{Kilpatrick23,Pledger23,Neustadt2024,VanDyk23,Xiang23, Jencson23,Soraisam23,Qin23,Niu23,Ransome2024}, UV spectroscopy \citep{Bostroem2024,Zimmerman23}, polarimetric observations \citep{Vasylyev23}, early X-ray \citep{Mereminiskiy2023-atel,Chandra23,Panjkov23,grefenstette2023} and radio observations \citep{Berger23,chandra2023-atel-radio,Matthews-atel16056,Matthews2023-Atel-detection}. 

Here, we present results from our extensive X-ray and radio monitoring campaign of \sn{} spanning $\delta t \approx 4-165$\,d. The paper is organized as follows. In \S\ref{sec:obs and data reduction}, we present observations and data reduction. Inferences from our multi-epoch broad-band X-ray spectral analysis are described in \S\ref{Sec:XrayInferences}. In \S\ref{Sec:RadioInferences}, we present the modeling of radio synchrotron emission and results. We discuss our findings in \S \ref{Sec:Disc} and conclude in \S\ref{Sec:conclusions}.
%%%%%%%%%%%%%%%%%%%%%%%%%%%%%%%%%%%%%%%%%%%%%%%%%%%%%%%%%
%%%%%%%%%%%%%%%%%%%%%%%%%%%%%%%%%%%%%%%%%%%%%%%%%%%%%%%%%%%%%%%%%
\section{Observations and Data reduction} \label{sec:obs and data reduction}
%--------------------------------------------------------------
\subsection{NOEMA} \label{SubSec:NOEMAdata}
We observed the field of \sn{} with the Northern Extended Millimetre Array (NOEMA) under Director's Discretionary Time (project ID d23ab; PI: T.\ Laskar) in June and July 2023 using the most compact configuration with baselines ranging between 17m and 175m. Observations took place over three epochs at 3mm (tuned to 89.9 GHz in the upper side band) and 2mm (tuned to 149.9 GHz in the upper side band) each, with $\leq$1day difference per epoch between the two bands. We used dual polarization and an initial spectral resolution of 2~MHz over the full bandwidth. We used the radio star MWC349 for absolute flux calibration for most tracks, except for one track during which we used the radio star LkHa101. The continuum fluxes of both sources are regularly monitored against planets, giving an uncertainty of about 5-10\% at 3mm and 10-15\% at 2mm. The bandpass calibration was done on one of the following sources: 2013+370, 3C273, 1418+546, 3C84, or 2200+420, while the temporal gains on phases and amplitudes were calibrated against 1418+546. Each epoch was observed on-source for about 1.1~h at 3mm and between 0.6~h and 2.1~h at 2mm. We calibrated the data with the GILDAS package CLIC, followed by imaging with the Common Astronomy Software Applications (CASA, version 6.5.5.1; \citealt{McMullin2007,CASAteam2022}) using natural weighting. We detect a point source in all three epochs at 3mm but not at 2mm. We measured the flux density by fitting a point source at the position of the optical counterpart. We report the details of our observations and flux measurements in Table~\ref{Tab:radio-NOEMA}. 
%---------------------------------------------------------------
\subsection{VLA} \label{SubSec:VLAdata}
We observed the field of \sn{} with the Karl G. Jansky Very Large Array (VLA) starting from 2023 May 23.01 ($\delta t = 4.2$\,d) under project code SF151070 (PI: R. Margutti). Our observations span a time period of $\delta t\approx 4 - 165$\,d covering frequencies 2$-$24 GHz with the VLA in A, BnA, and D configurations. We used 3C286 as the flux density and bandpass calibrator and J1419+5423 as the phase calibrator. The data were calibrated using the VLA calibration pipeline in CASA version 6.4.1.12 and imaged using task \texttt{TCLEAN} \citep{Offringa2017} adopting natural weighting. We performed phase-only self-calibration where necessary.

\begin{figure} 
 	\centering
 	\includegraphics[width=0.48\textwidth]{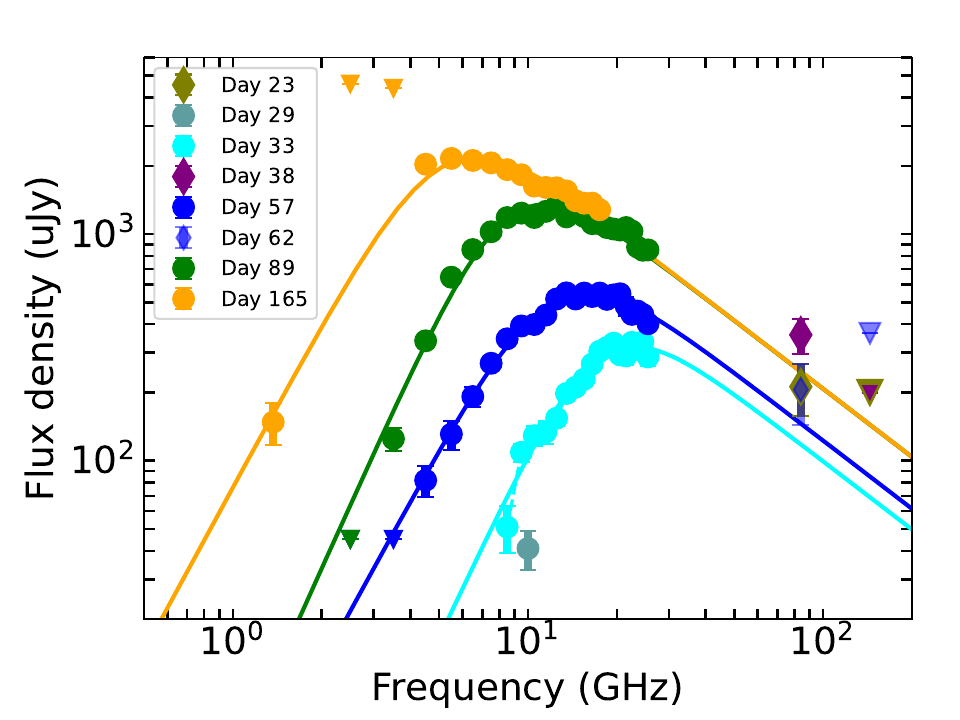} 
  \caption{Radio spectra of \sn{} in the time range $\delta t = 23-165$\,d. The flux density measurements at 84 GHz obtained using NOEMA are marked as diamonds. Inverted triangles mark the 5$\sigma$ flux density upper limits. Solid lines represent best-fit broken power-law models with smoothing parameter fixed to $s = -1$ and optically thin spectral index fixed to $\alpha_{1} = -1$.}
 \label{Fig:radioSED-fit}
 \end{figure}

Radio emission was not detected in our observation on 2023 May 23.01 ($\delta t=4.2$\,d) with a 5$\sigma$ flux density limit of $F_{\nu}<33\,\mu$Jy \citep{Matthews-atel16056}. We detected radio emission from the SN on 2023 Jun 17.04 ($\delta t=29.2$\,d) with a flux density of $F_{\nu}=(41\pm8)\,\mu$Jy at 10 GHz \citep{Matthews2023-Atel-detection}. The flux density and position of the emission were measured using task \texttt{IMFIT} that fits an elliptical Gaussian of the size of the synthesized beam at the source position in the image plane. The SN was detected at multiple frequencies in the successive epochs of observations. We present the flux densities in Table \ref{AppendixRadio} and spectra in Figure \ref{Fig:radioSED-fit}. The VLA was in D configuration on 2023 Oct 31 ($\delta t\approx165$\,d), and the SN emission was unresolved with other nearby sources in the 2$-$5 GHz maps, resulting in relatively high flux density limits.  
%---------------------------------------------------------------
\subsection{GMRT} \label{SubSec:GMRTdata}
We observed the field of \sn{} with the Giant Metrewave Radio Telescope (GMRT) from 2023 May 21 ($\delta t\approx2$\,d) to 2023 Oct 31 ($\delta t\approx165$\,d) under project code 44$_{-}$095 and 45$_{-}$091 (PI: Poonam Chandra). The data were recorded in standard continuum mode in bands 4 (550$-$850 MHz) and 5 (1050$-$1450 MHz) with a bandwidth of 400 MHz split into 2048 channels. 3C286 was used as the flux density and bandpass calibrator, whereas J1400+621 was used as the phase calibrator. We processed the raw visibilities using Astronomical Image Processing Software \citep[AIPS;][]{Greisen2003} following standard procedure \citep{Nayana2017}. The data were initially flagged and calibrated using various AIPS tasks. The fully calibrated target source data were imaged using task \texttt{IMAGR}. We performed a few rounds of phase-only self-calibration to improve the image fidelity.

Radio emission was not detected until 2023 Oct 30 ($\delta t\approx165$\,d) resulting a 3$\sigma$ flux density upper limit $F_{\nu}<75\,\mu$Jy at 1.25\,GHz \citep{chandra2023-atel-radio} and $F_{\nu}<90\,\mu$Jy at 0.65\,GHz. We detected radio emission from the SN with a flux density $F_{\nu}=(148\pm31)\,\mu$Jy at 1.25 GHz on 2023 Oct 30 ($\delta t\approx165$\,d). The flux density was measured by fitting an elliptical Gaussian at the SN position in the image plane using task \texttt{JMFIT}. Details of GMRT observations and flux measurements are reported in Table \ref{Tab:radio-gmrt}.

\begin{figure*} 
    \centering
    \includegraphics[width=1\textwidth]{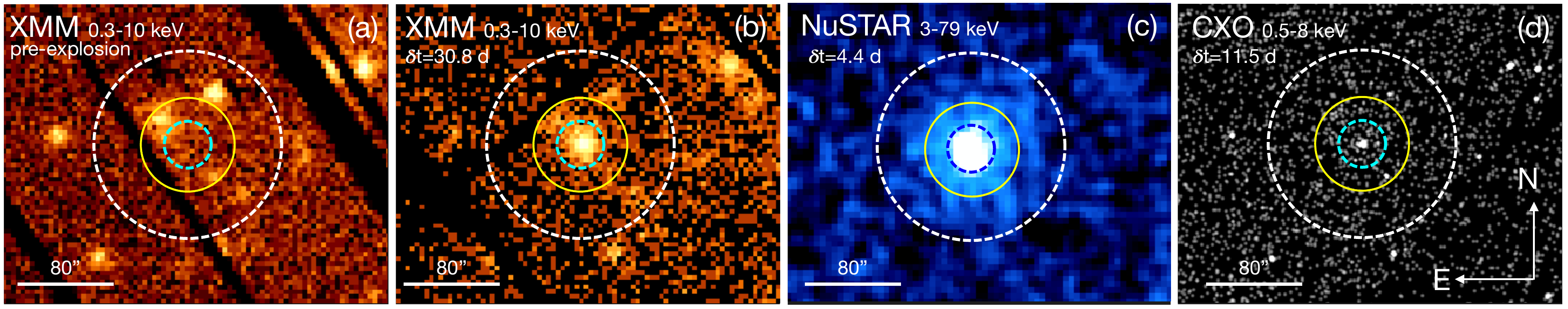}
     \caption{Pre-explosion (panel \emph{(a)}, obs ID 0824450501, exposure of $\sim55$\,ks) and post-explosion (panel \emph{(b)}, obs ID 0921180101, exposure of 10\,ks) \xmm/EPIC-pn soft-X-ray images of the field of \sn{}, showing the presence of three unrelated sources within $40\arcsec$ (yellow region) that contaminate the \xmm\, and Swift-\emph{XRT} PSF. Panel \emph{(c)} \nustar\, hard X-ray image of the field (obs ID 90902520002, exposure of $\sim42\,$ks). 
     Panel \emph{(d)} Thanks to the significantly sharper PSF,  contamination by unrelated sources is not an issue for \chandra\, observations of \sn{}\, (obs ID 27862, exposure of $\sim20$\,ks). Dashed cyan or blue region: $20\arcsec$ radius region at the location of \sn{}. Yellow region: $40\arcsec$ radius region (which is representative of the \xmm\, and \emph{XRT} PSF). The dashed white region has a radius of $80\arcsec$.}
 \label{Fig:Xraycontamination}
 \end{figure*}

%---------------------------------------------------------------
\subsection{Swift-XRT}  \label{SubSec:XRTdata}
\begin{figure} 
 	\centering
 	\includegraphics[width=0.48\textwidth]{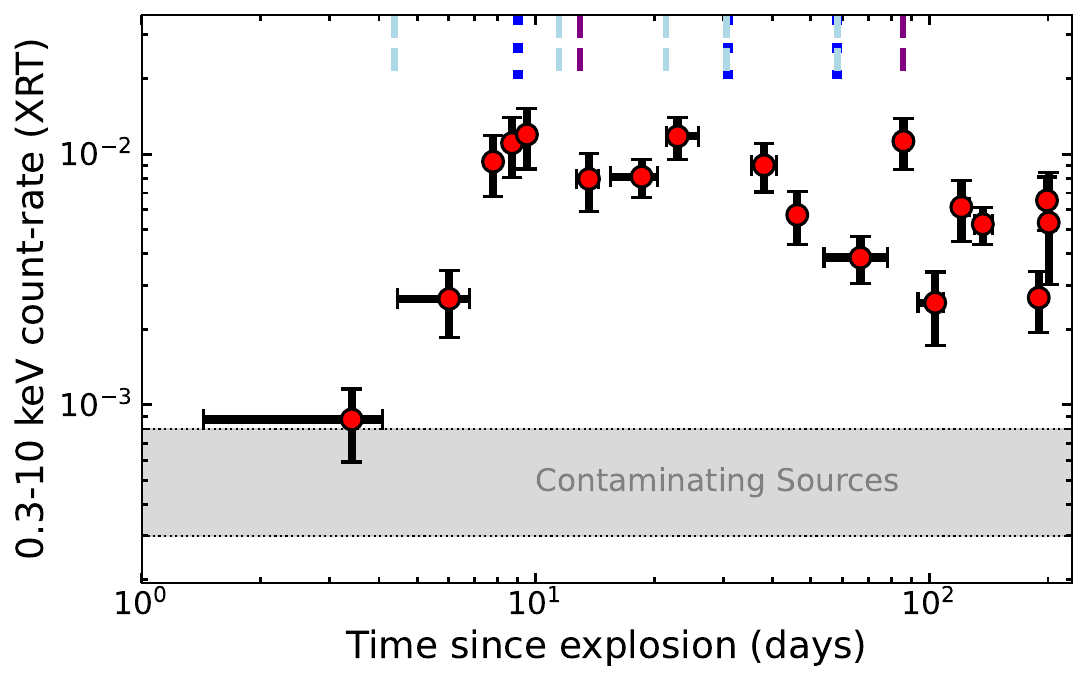}
         \caption{Post-explosion \emph{Swift}-XRT count-rate light curve. Vertical lines mark the time of \xmm\, (dark blue, dotted), \nustar\, (light-blue, dashed), and \chandra\,(purple, dashed) observations (some epochs are coordinated). The grey area marks the range of the pre-explosion X-ray emission due to contaminating sources as detected by \emph{Swift}-XRT. The observed steep count-rate increase at early times mostly results from the rapidly decreasing intrinsic hydrogen column density with time accompanied by cooling (Fig. \ref{Fig:TemperatureNH}). }
 \label{Fig:XrayGen}
 \end{figure}

The X-Ray Telescope (XRT, \citealt{Burrows05}) onboard the Neil Gehrels Swift Observatory \citep{Gehrels04} started observing \sn{} on 2023-05-20 UT06:25:01 ($\delta t\approx 1.4$\,d). Here we present the analysis of XRT observations acquired until  2023-10-18. Data have been processed with HEASoft v6.33 and corresponding calibration files. 
We extracted a 0.3--10 keV count-rate light-curve and rebinned to minimum count-rate following standard procedures (\citealt{Evans09,Margutti13}). We extracted 12 post-explosion spectra at salient phases of the SN X-ray light-curve, and around the time of acquisition of \nustar\, observations (Table~\ref{Tab:Xcontimuum}). The \emph{Swift}-XRT and the broad-band (soft and hard X-ray) spectral fitting is described in \S\ref{SubSec:X-raySpec}. 

\emph{Swift}-XRT serendipitously observed the location of \sn{} in a number of occasions starting $\approx6700$\,d before explosion. X-ray emission is detected at the location of \sn{} at the level of $\approx 4\times 10^{-4}\,\rm{c\,s^{-1}}$ (0.3--10 keV), which corresponds to a flux in the range $\approx (5-10)\times 10^{-15}\,\rm{erg\, s^{-1}cm^{-2}}$ in the very soft energy band 0.3--1 keV for an assumed power-law model with photon index in the range $\Gamma=2-4$. This emission is physically unrelated to \sn{}. The X-ray spectrum at very soft energies ($<1\,$keV) at early times ($\delta t<10$\,d)  when the \sn{} emission is heavily absorbed is the most affected by the contaminating emission. \xmm\, and \chandra\,observations (\S\ref{SubSec:CXOdata}-\S\ref{SubSec:XMMdata}) reveal that this ``host-galaxy background'' flux mostly results from the presence of three X-ray sources within $\approx 40\arcsec$ of \sn{} that contaminate the \emph{Swift}-XRT Point Spread Function (PSF). Consistent with the conclusion of \cite{Panjkov23}, we ascribe the observed time variability of the X-ray emission in pre-explosion images to the contamination by physically-unrelated nearby sources.  

X-ray emission at the location of \sn{} is detected starting at around $\approx 4$\,d and we show the \emph{Swift}-XRT count-rate light-curve of \sn{} in Fig. \ref{Fig:XrayGen}. Accurate time-dependent flux calibration is performed as described in \S\ref{SubSec:X-raySpec}. We note that we quantitatively accounted for the contaminating flux using the spectral model derived from pre-explosion \xmm\, observations in \S\ref{SubSec:XMMdata}.

%---------------------------------------------------------------
\subsection{Chandra X-ray Observatory (CXO)} \label{SubSec:CXOdata}
The \chandra\,started observing \sn{} on 2023-05-31 (obsID 27862, $\delta t=12.84$\,d) for an exposure time of 20.1\,ks, with a second epoch of \chandra\,observations acquired starting on 2023-08-11  ($\delta t\approx 85$\,d, obsIDs 27863 and 28374 for a total exposure of 20.9\,ks, PI Chandra). \chandra\,ACIS-S data have been reduced following standard practice within \texttt{CIAO v4.15} and corresponding calibration files. A bright X-ray source is detected with high confidence at the location of \sn{} %, with net count-rates reported in 
(Table \ref{Tab:Xraylog}). The detection significance was estimated with \texttt{wavdetect}. We extracted a spectrum with \texttt{specextract} using a source region with a radius of $2\arcsec$ (first epoch) and $1.5\arcsec$ (second epoch). The background was estimated from a source-free region with a radius of $33.5\arcsec$. While extensive \chandra\,pre-explosion observations at the \sn{} location exist (and no evidence for X-ray emission has been found, \citealt{grefenstette2023}), the sharp PSF of the \chandra\,implies that contamination by unrelated sources is not an issue for \chandra\,observations of \sn{}. Spectral modeling is described in \S\ref{SubSec:X-raySpec}.

%----------------------------------------------------------------------------------
\subsection{XMM-Newton} \label{SubSec:XMMdata}
\begin{figure*} 
    \centering
    \includegraphics[width=1.\textwidth]{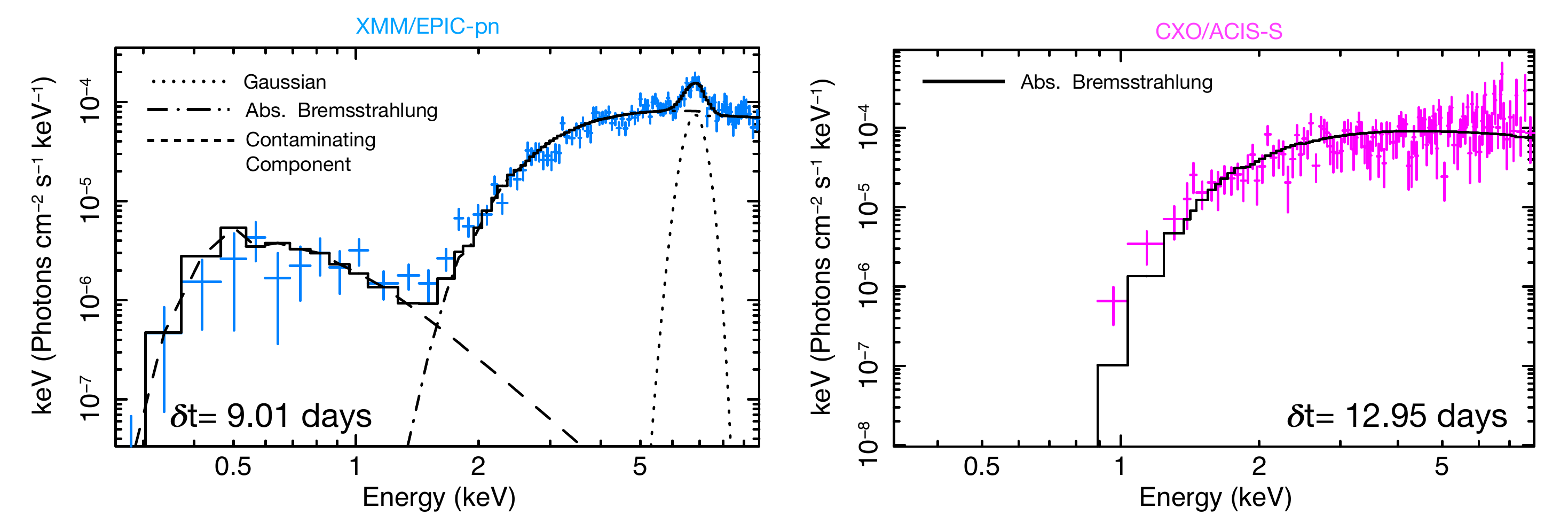}
     \caption{ \emph{Left Panel:} \xmm\, EPIC-pn unfolded $F_{\nu}$ spectrum at $\delta t=9.01\,$d showing a clear evidence for an excess of soft X-ray emission at $<1$\,keV with respect to an absorbed thermal bremsstrahlung model. A Gaussian profile has been added to model the Fe line emission at $\approx 6.8\,$keV.  \emph{Right Panel:} \chandra\,ACIS-S unfolded spectrum of \sn{} at  $\delta t=12.95\,$d with best-fitting absorbed thermal bremsstrahlung model. Since (i) the soft X-ray component in the \xmm\, spectrum has properties consistent with those inferred from pre-explosion observations; and (ii) \chandra\,observations with significantly sharper PSF do not require the addition of a soft-X-ray component, we conclude that there is no compelling observational evidence that the soft X-ray excess of emission is physically related to \sn{} and we thus ascribe that to contamination by unrelated sources in the host galaxy (Fig.\ref{Fig:Xraycontamination}).  }
 \label{Fig:Xraycontaminationspec}
 \end{figure*}

\begin{figure*} 
    \centering
    \includegraphics[width=1.\textwidth]{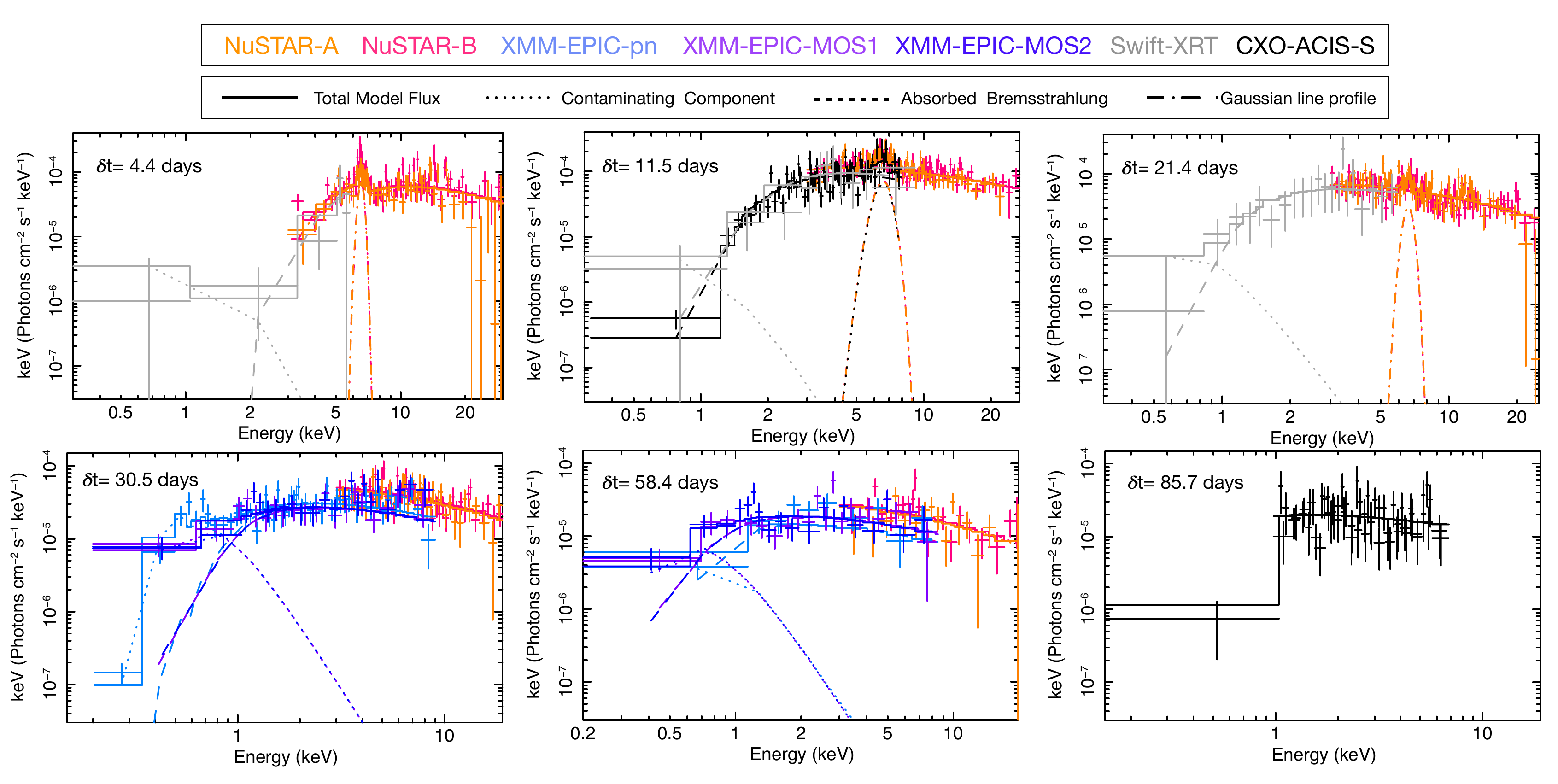}
     \caption{Unfolded $F_{\nu}$ broad-band X-ray spectra for the six epochs for which we report the best fitting parameters in Table~\ref{Tab:Xcontimuum}. In all cases the data are best fit by an absorbed thermal bremsstrahlung model (long dashed lines) with a decreasing level of intrinsic absorption with time. Dotted lines: contamination component from unrelated sources in the host galaxy. Fe emission, here modeled with a Gaussian profile (dot-dashed line), is clearly detected in the first three epochs. Data have been rebinned for displaying purposes only. \chandra\,data below 0.5 keV are here shown for comparison but have not been used for fitting.  }
 \label{Fig:XraySpectraPlot}
 \end{figure*}

A sequence of three deep \xmm\, observations of \sn{} have been acquired in the time range $\delta t\approx 9.0-58.2$\,d (PIs Campana, Margutti). Details are reported in Table \ref{Tab:Xraylog}. We reduced and analyzed the data of the three European Photon Imaging Camera (EPIC)-pn, MOS1, and MOS2 using the Scientific Analysis System (SAS) v.20.0.0 and corresponding calibration files (CALDB 3.13). We filtered out the intervals of time affected by severe proton flaring with the {\tt\string espfilt\footnote{\url{https://xmm-tools.cosmos.esa.int/external/sas/current/doc/espfilt/espfilt.html}}} task, which led to a significant reduction of the effective exposure time of the third \xmm\, observation (Table~\ref{Tab:Xraylog}). A bright X-ray source is detected in all observations at the location of \sn{} with count-rates reported in Table \ref{Tab:Xraylog}. To assess the significance of the detection, we ran the task {\tt\string edetect\_chain}. For each of the three \xmm\, epochs we extracted three source spectra (i.e., one for the EPIC-pn, MOS1, and MOS2) using a region with $20\arcsec$ radius to minimize the impact of the contamination by nearby X-ray sources (Figure~\ref{Fig:Xraycontamination}). 

To further quantitatively assess the level of contamination of post-explosion \xmm\, observations of \sn{} by physically unrelated sources, we used a deep \xmm\, pre-explosion observation of the SN field acquired in 2018 (ObsID 0824450501, PI G. Israel, effective exposure time after accounting for proton flaring of 55.5 ks). Significant X-ray emission with a very soft spectrum is detected at energies $\lesssim 5\,\rm{keV}$ in a $20\arcsec$ region centered at the SN location. The EPIC-pn spectrum is well fit by a simple-power law model (\texttt{tbabs*pow}, phenomenological model) with best-fitting photon index $\Gamma =5.5^{+1.0}_{-0.6}$ and hydrogen absorption column density $\rm{NH_{pre}}=(0.44^{+0.14}_{-0.11})\times 10^{22}\,\rm{cm^{-2}}$. The corresponding observed flux in the 0.3--1 keV energy range is $F_{\rm{X,pre}}=(6.4^{+0.4}_{-0.6})\times 10^{-15}\,\rm{erg\,s^{-1} cm^{-2}}$. This flux estimate is consistent with our inferences based on pre-explosion \emph{Swift}-XRT observations (\S\ref{SubSec:XRTdata}). It follows that the contamination by nearby sources can have a non-negligible effect either (i) at early times $\delta t<9\,$d when the X-ray emission of \sn{} is highly absorbed, and the contaminating flux can dominate the observed emission in the very soft (e.g., 0.3--1 keV) energy range; or (ii) at very late times when the \sn{} emission becomes faint and eventually comparable to the astrophysical background. 

Case (i) applies to the first \xmm\, epoch, and to early \emph{Swift}-XRT observations acquired at $\delta t\lesssim 12$\,d (i.e. during the rise-time of the observed light-curve). We show in Figure \ref{Fig:Xraycontaminationspec} (left panel) the first \xmm\, spectrum where the presence of an excess of very soft X-ray emission at energies $\lesssim 1$\,keV compared to the absorbed thermal bremsstrahlung model is apparent. Modeling the excess of emission with an absorbed power-law model leads to parameter inferences that are statistically consistent  with those derived from the pre-explosion spectrum (with a 0.3--1 keV flux of $F_X\approx 2\times 10^{-15}\,\rm{erg\,s^{-1}cm^{-2}}$, even lower than in pre-explosion images). \chandra\, observations acquired $\approx 4$\,d later do not provide evidence for the presence of a soft X-ray excess. Taken at  face value, the excess of soft X-ray emission can also be explained by a partial covering fraction of the absorber.  However, given the consistency of the soft X-ray component detected by \xmm\, with pre-existing emission at the location of \sn{}, and the lack of evidence for this component in \chandra\,data, in the following we proceed under the hypothesis that the soft X-ray excess of emission is physically unrelated to \sn{} and more likely associated with emission of nearby X-ray sources that contaminate the wider \xmm\, and \emph{Swift}-XRT PSF. \chandra\,and \nustar\, observations are unaffected because of the significantly sharper PSF (\chandra), and the harder energy response (\nustar).

%----------------------------------------------------------------------------------
\subsection{NuSTAR (3--79 keV)} \label{SubSec:NuSTARdata}
\cite{grefenstette2023} presented the analysis of the first two epochs of \nustar\, data. Here we present the homogeneous analysis of the entire \nustar\, data set of \sn{}\,, which comprises five epochs of observations acquired between $\delta t\approx 4.3-58.4$\,d (PIs Grefenstette, Margutti). Details are provided in Table~\ref{Tab:Xraylog}. 
We extracted spectra and response files using the NuSTAR Data Analysis Software (v2.1.2) and calibration files (version 20240104), with \texttt{nupipeline} and \texttt{nuproducts}. For the source, we used an extraction region with a radius varying from $80\arcsec$ (at maximum source brightness) to $50\arcsec$  (at minimum source brightness) to maximize the S/N ratio. A nearby source-free region was used to estimate the local background. We checked for the presence of solar flares and significant radiation belt backgrounds using the standard background plots. We did not find any, so we used the standard nustardas filtering.
In the following, we thus proceed with data products from the standard \texttt{nupipeline} extraction, which benefits from the larger number of counts. Spectral modeling is done jointly with the soft X-ray data in \S\ref{SubSec:X-raySpec}.

\startlongtable
\begin{deluxetable*}{clccccc}
\tablecaption{Best-fitting parameters of the absorbed thermal bremsstrahlung continuum model for the five epochs with broad-band X-ray coverage, and for the last epoch of CXO observations.  }
\tablehead{
\colhead{Time} &\colhead{Instrument} & \colhead{$\rm{NH_{\rm{int}}}^{a}$} & \colhead{$T$} &\colhead{Log(Flux$^{b})$} &\colhead{Log(Flux$^{b})$}& \colhead{Norm$^{c}$}\\
 (d) &  & ($10^{22}\,\rm{cm^{-2}}$) & (keV) & 0.3-10 keV & 0.3-30 keV & ($10^{-4}$)}
\startdata
4.4 &   NuSTAR+XRT      &$31.2^{+4.3}_{-7.6}$&$41.3^{+58.6}_{-9.8}$& $-11.83^{+0.04}_{-0.08}$ &$-11.47^{+0.04}_{-0.03}$ &$3.55^{+0.20}_{-0.28}$\\
11.5 &  NuSTAR+XRT+CXO  &$3.2^{+0.3}_{-0.3}$&$61.5^{+22.9}_{-9.4}$& $-11.77^{+0.01}_{-0.02}$&$-11.42^{+0.02}_{-0.01}$&$4.29^{+0.19}_{-0.12}$\\
21.4 &  NuSTAR+XRT      &$1.5^{+0.7}_{-0.3}$&$31.6^{+6.8}_{-4.8}$&  $-12.08^{+0.02}_{-0.02}$&$-11.77^{+0.02}_{-0.02}$&$1.97^{+0.05}_{-0.06}$\\
30.5 &  NuSTAR+XMM      &$0.63^{+0.10}_{-0.12}$& $31.4^{+13.0}_{-4.4}$& $-12.23^{+0.01}_{-0.03}$&$-11.93^{+0.03}_{-0.02}$&$1.38^{+0.06}_{-0.07}$ \\
58.4 &  NuSTAR+XMM      &$0.34^{+0.10}_{-0.09}$&$22.4^{+10.4}_{-5.8}$& $-12.52^{+0.02}_{-0.06}$&$-12.25^{+0.05}_{-0.03}$&$0.70^{+0.05}_{-0.07}$ \\
85.7 & CXO & $0.06^{+0.33}_{-0.01}$ & $>7$& $-12.58^{+0.06}_{-0.02}$ & $-12.55^{+0.074}_{-0.075}$ & $0.54^{+0.18}_{-0.04}$\\
\enddata
\tablecomments{$^{a}$\,Solar abundance assumed. $^{b}$\,Unabsorbed, units of $\rm{erg\,s^{-1}cm^{-2}}$. $^{c}$\,Normalization of the bremsstrahlung model defined as $\rm{Norm}\equiv\frac{3.02\times 10^{-15}}{4\pi d^2  }\int  n_e n_I dV$, where $n_e/\rm{cm^{-3}}$ and $n_I/\rm{cm^{-3}}$  are the electron and ion number densities, respectively, and $d/\rm{cm}$ is the distance to the target.
\label{Tab:Xcontimuum}}
\end{deluxetable*}

%\startlongtable
\begin{deluxetable}{cccc}
\tablecaption{Best-fitting parameters of the Fe emission modeled with a Gaussian profile centered at $E_{0}$ with standard deviation $\sigma$.}
\tablehead{
\colhead{Time} &\colhead{$E_{0}$} & \colhead{$\sigma$} & \colhead{Log(Flux$^{a})$} \\
 (d) &(keV)  & (keV) & }
\startdata
4.4 & $6.54^{+0.06}_{-0.08}$ & $0.2^{b}$ & $-13.06^{+0.07}_{-0.10}$\\
11.5 & $6.52^{+0.07}_{-0.11}$ & $0.58^{+0.20}_{-0.07}$ & $-12.77^{+0.09}_{-0.03}$\\
21.4 & $6.61^{+0.14}_{-0.12}$& $0.33^{+0.22}_{-0.11}$& $-13.39^{+0.11}_{-0.19}$\\
\enddata
\tablecomments{$^{a}$\,Unabsorbed, 0.3-10 keV, units of $\rm{erg\,s^{-1}cm^{-2}}$. $^{b}$\,Set by the instrumental resolution.
\label{Tab:FeLine}}
\end{deluxetable}

%---------------------------------------------------------------
\begin{figure} 
 	\centering
 	\includegraphics[width=0.48\textwidth]{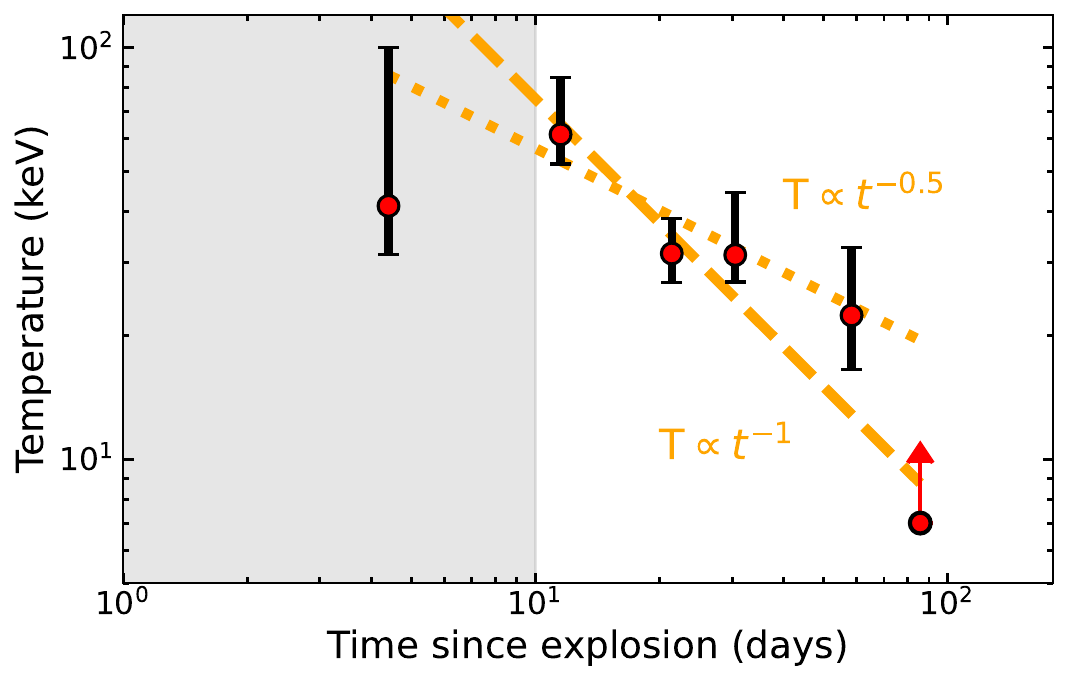}
   	\includegraphics[width=0.48\textwidth]{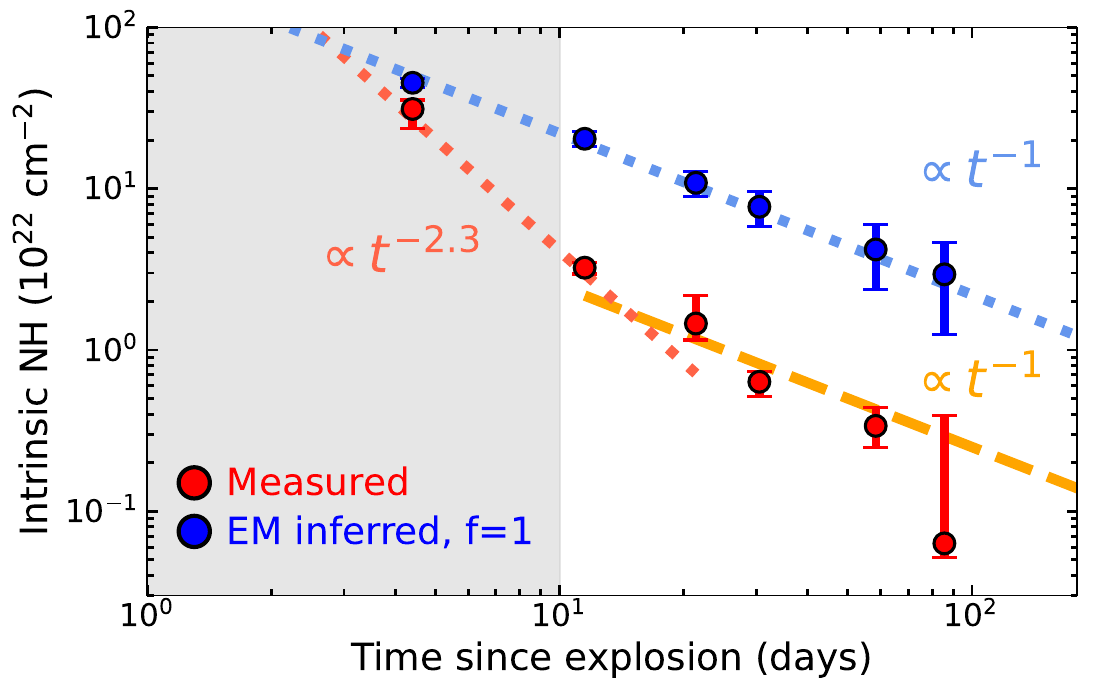}
   	\includegraphics[width=0.50\textwidth]{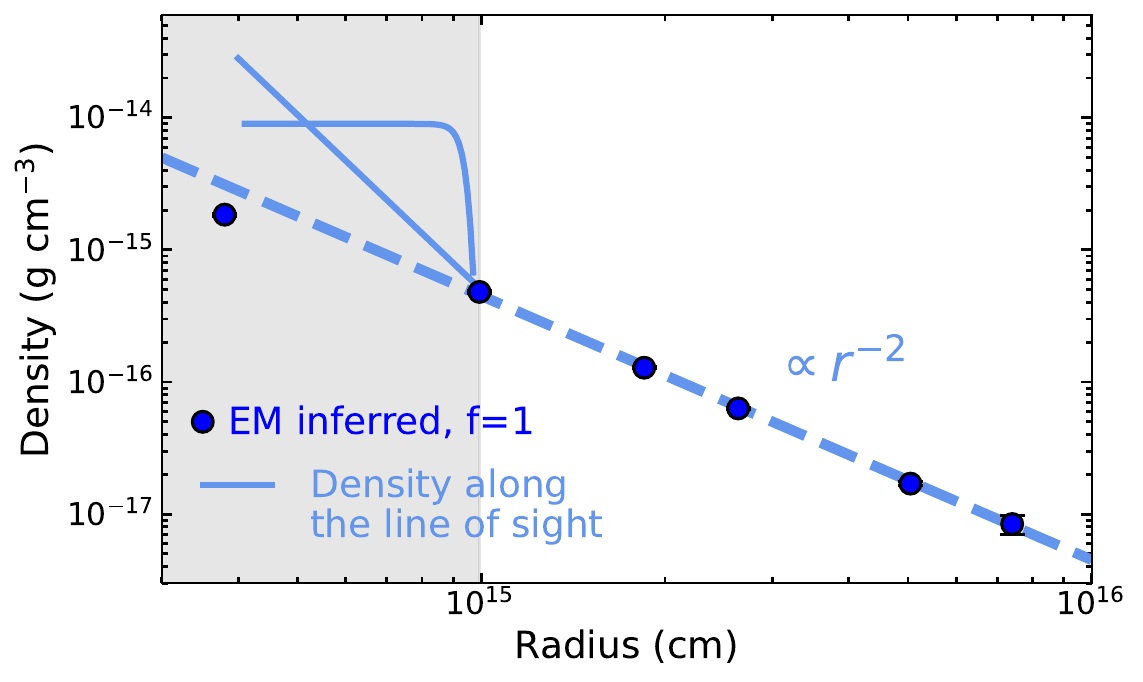}
       \caption{Evolution of the X-ray emitting %plasma
       electron temperature (\emph{upper panel}), intrinsic neutral hydrogen column density (\emph{middle panel}), and density (\emph{lower panel}) as inferred from the modeling of the broad-band observations (\S\ref{SubSec:X-raySpec}).  
       The dashed and dotted lines have been added to guide the eye. For comparison, $\rm{NH_{int}}\propto t^{-1}$ is expected for a shock evolution in a wind-density profile $\rho_{\rm{CSM}}\propto r^{-2}$ with constant neutral fraction and an emitting radius $R_{sh}\propto t$. The absorption column of a radiatively cooling shell in the RS would also evolve $\propto t^{-1}$ for a constant mass-loss rate scenario (e.g., \citealt{Nymark06}). The grey area at $\delta t<10$\,d (or $r<10^{15}\,\rm{cm}$) highlights the period of strong interaction (i.e., narrow lines present in the optical spectra). The EM-inferred $NH_{\rm{int}}$ and density (blue filled circles) assume a filling factor $f=1$. The blue solid lines in the lower panel show two examples of density profiles along our line of sight that satisfy the observational constraints of \S\ref{Sec:XrayInferences}. } 
 \label{Fig:TemperatureNH}
 \end{figure}

\begin{figure} 
 	\centering
 	\includegraphics[width=0.45\textwidth]{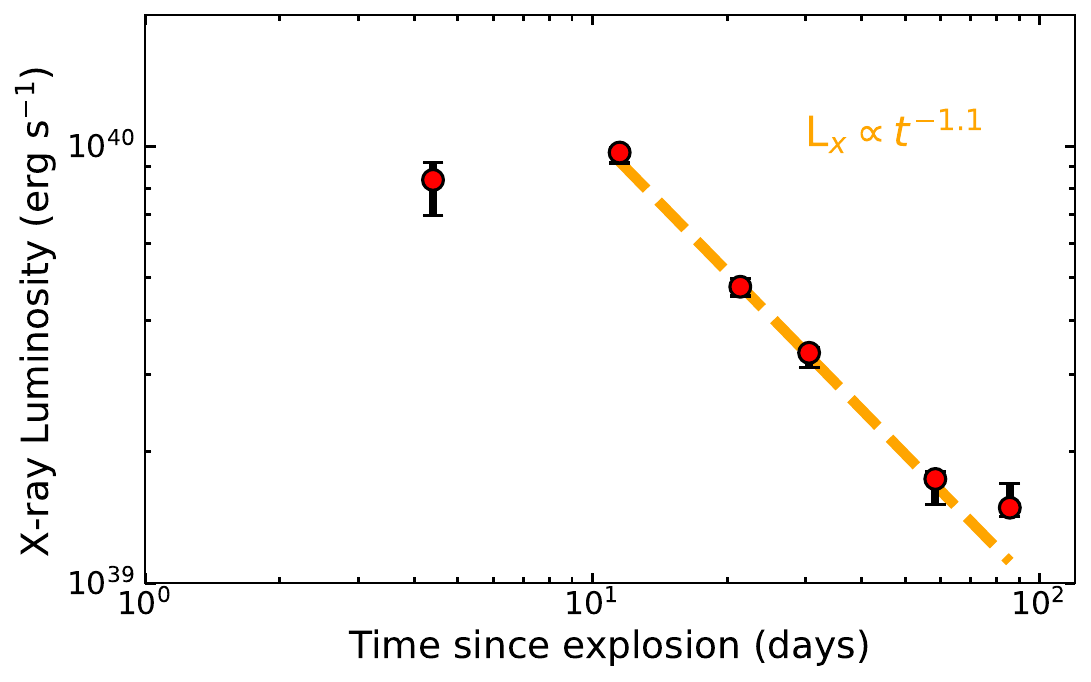}
     \caption{\emph{Unabsorbed} 0.3-10 keV X-ray luminosity of \sn{}\, (red points) estimated using the broad-band X-ray spectral modeling of \S\ref{SubSec:X-raySpec} that properly accounts for the spectral evolution of the source with time. We find that $L_X$ is roughly constant at $\delta t\lesssim 10$\,d, (i.e., before the emergence of broad UV spectral features, \citealt{Zimmerman23}), and later decays as $L_X\propto t^{-1.1}$ (orange dashed line). 
     }
 \label{Fig:XrayLumIC}
 \end{figure}
 
\subsection{Joint soft and hard X-ray spectral modeling and X-ray flux calibration} \label{SubSec:X-raySpec}
Broad-band X-ray spectral modeling is necessary to reliably estimate the spectral parameters and to constrain the physical properties of the emitting region. Below, we first perform joint soft and hard X-ray spectral modeling for the five epochs in the time range $\delta t=4.4-58.4$\,d for which \sn{} was observed with \nustar\, in coordination with soft X-ray telescopes (\emph{Swift}-XRT, \xmm\, or \chandra). We then use the spectral parameters from this broad-band analysis to anchor the flux calibration of the intermediate epochs when only soft X-ray data are available.

As observationally confirmed by the extensive campaign of SN\,2014C \citep{Margutti17,Brethauer22,Thomas22}, the X-ray emission from SN shocks that interact with dense material is expected to be (i) highly absorbed, (ii) with a level of absorption that decreases with time, and (iii) of thermal nature. \sn{} broadly confirms these expectations and also shows an excess of emission around $\sim6.5$\,keV that has been attributed to Fe K$\alpha$ lines in strongly interacting SNe such as 2014C.  We use a Galactic neutral hydrogen column density in the direction of \sn{}\, of $\rm{NH_{MW}}=7.9\times 10^{20}\,\rm{cm^{-2}}$ \citep{HI4PI}.\footnote{Adopting the lower value  of $\approx 2\times 10^{20}\,\rm{cm^{-2}}$ inferred by \cite{Willingale13} based on the X-ray afterglow of Gamma-Ray Bursts has no impact on our major conclusions.}
We adopt the Solar abundances as updated in \cite{Asplund09} (i.e., \texttt{abund aspl} within \texttt{Xspec}\footnote{We note that using the default Solar abundances in \texttt{Xspec} (i.e. those from \citealt{Anders89}) results in NH estimates that are a factor $\sim2$ different.}). We first use \texttt{apec} to simulate the spectrum of a collisionally-ionized plasma (continuum and corresponding line emission). However, a single-temperature component \texttt{apec} model fails to simultaneously reproduce the bright Fe line emission \emph{and} the underlying continuum, as expected. We comment on the implications of this finding in \S\ref{Sec:XrayInferences}. The maximum temperature  $T=64$\,keV allowed within the \texttt{apec} model is also not physical for this case. We thus proceed with an absorbed thermal bremsstrahlung model (\texttt{tbabs*ztbabs*bremss} within \texttt{Xspec}) for the continuum, and we model the excess of emission with a Gaussian line profile. For \xmm\, (EPIC-pn, MOS1, and MOS2) and \emph{Swift}-XRT observations, we model the contamination by nearby X-ray sources adopting as a phenomenological model an absorbed power-law  (\texttt{tbabs*pow}) with photon index and neutral hydrogen column densities frozen to the best-fitting values obtained from deep pre-explosion \xmm\, observations (\S\ref{SubSec:XMMdata}) and normalization free to vary to account for potential time-variability of the emission from the contaminating sources when needed (as observed by \emph{Swift}-XRT, \S\ref{SubSec:XRTdata}; see also \citealt{Panjkov23}). Finally, we account for uncertainties affecting the instrumental cross calibration in the absolute flux phase space by adding a multiplicative factor to the model.  Absolute fluxes are reported with respect to \nustar\, module A. We employ the W statistic and we self-consistently derive the parameter uncertainties with MCMC. The best-fitting model parameters are listed in Table \ref{Tab:Xcontimuum} and \ref{Tab:FeLine} for the bremsstrahlung continuum and the emission line profile, respectively. 

We model the \emph{Swift}-XRT, \xmm\, and \chandra\, spectra without a \nustar\, counterpart using an absorbed bremsstrahlung spectrum with temperature and neutral hydrogen absorption fixed at the values obtained by interpolating those derived from the joint spectral analysis above. The results from this analysis are then used to derive a time-dependent count-to-flux (absorbed and unabsorbed) conversion factor. Our 0.3--10 keV and 0.3--30 keV flux-calibrated light-curve data are reported in Table \ref{Tab:Xcontimuum}. The last epoch of \chandra\,observations acquired at $\delta t=85.7$\,d constraints the plasma temperature to values $T>7$\,keV and it is consistent with minimal intrinsic absorption $\rm{NH_{int}}\approx 0.06\times 10^{22}\,\rm{cm^{-2}}$ (albeit we note that the upper bound of this parameter is poorly constrained).
%, with values consistent with those of the previous epoch.

We end with a comparison with results from the analysis of partial X-ray data sets of \sn{} that have appeared in the literature. First, by performing a time-resolved soft and hard X-ray joint spectral fitting, we do not confirm the findings of a two-temperature continuum by \cite{Panjkov23}. We ascribe the difference to their method of fitting a time-averaged spectrum of a source that is strongly temporally evolving both in terms of absorption column and temperature of emission (while also not accounting for the soft contamination component from the host galaxy). Second, while the \emph{observed} (i.e., absorbed) 0.3--10 keV flux (or count-rate) light-curve of \sn{}\, shows a marked rise with time until $\delta t\approx 10$\,d to then ``plateau'' until $\delta t\approx 200$\,d  (Fig. \ref{Fig:XrayGen}), we find that this apparent behavior is due to the combined evolution of $T(t)$ and $\rm{NH_{int}}(t)$ to lower values and does \emph{not} represent the intrinsic evolution of the X-ray luminosity from \sn{}. Differently from \cite{Panjkov23,Zimmerman23}, our time-resolved broad-band X-ray spectral analysis indicates that the intrinsic broad-band 0.3--10 keV luminosity of \sn{}\, is $L_X\approx 10^{40}\,\rm{erg\, s^{-1}}$ for $\delta t \approx 4-10$\,d and shows a clear evidence for a temporal decay as $L_X\propto t^{-1.1}$ at $\delta t\gtrsim 10$\,d (Fig. \ref{Fig:XrayLumIC}). While based on more extensive broad-band X-ray observations, our X-ray modeling is overall consistent with \cite{Chandra23}. 
Finally, we confirm the evolution towards larger values of the Fe line width with time, as was reported by \cite{grefenstette2023}, which most likely results from the emergence of Fe emission with different stages of ionization (i.e., from mostly neutral to higher levels of ionization). 

%%%%%%%%%%%%%%%%%%%%%%%%%%%%%%%%%%%%%%%%%%%%%%%%%%%%%%%%%%%%%%%%%
\section{Inferences from X-ray observations}\label{Sec:XrayInferences}
The broad-band X-ray spectrum of \sn{}\, observed starting from day $\approx$\,4 is of thermal origin, and there is no evidence for a dominant non-thermal component at any time. The \emph{observed} soft X-ray flux (or count-rate) rises as a consequence of the decreasing absorption with time, which results from %the combination of 
the shock expansion through the medium and, possibly, from the larger ionization of the CSM with time. The intrinsic $L_X$ remains roughly constant until $\delta t\approx 10$\,d and later decays\footnote{The actual index depends on the X-ray energy band considered. Here we quote the value for 0.3--10 keV.} as $L_X\propto t^{-1.1}$.

The shock velocity determines the energy imparted to particles crossing it. It follows that the post-shock plasma temperature for the FS and the RS are expected to be different, giving an indication of which shock is dominating the emission. The large \emph{electron} temperatures $k_bT_e\ge 50$\,keV ($T_e>6\times 10^8\,\rm{K}$) inferred in the first epoch at $\approx 10$\,d (Table \ref{Tab:Xcontimuum}, Fig. \ref{Fig:TemperatureNH}) implies that the emitting region is the forward shocked CSM. The ion temperature of the shocked CSM gas is:  
\begin{equation}
\label{Eq:TiFS}
  T_{i, \rm FS}\approx 1.36\times 10^{9}\Big (\frac{n-3}{n-s} \Big)^2 \Big(\frac{v_{\rm sh}}{10^4\,\rm{km\,s^{-1}}} \Big )^2\, K
\end{equation}
(e.g., \citealt{Fransson96}, their Eq. 3.1), where $n$ parameterizes the outer ejecta density profile $\rho_{\rm ej}(r)\propto r^{-n}$, the CSM density scales as $\rho_{\rm{CSM}}\propto r^{-s}$, and we have assumed Solar abundances. For $n=12$, as appropriate for the convective envelope of a RSG star \citep{Matzner99,Ro13}, and a wind-density profile (\S\ref{SubSec:densityXrays}), $k_bT_{i, \rm FS}\approx 100\,$keV. For the same choice of parameters, shock jump conditions predict a reverse shock temperature that is a factor $\approx 80$ lower: 
\begin{equation}
\label{Eq:TiRS}
T_{i,\rm RS}=\frac{T_{i,\rm FS}}{(n-3)^2},
\end{equation}
which implies $T_{i,\rm RS}\sim$\, 0.5 keV. This value is consistent with shock temperatures found in other non-strongly interacting Type IIP SNe (e.g., SNe\,2004dj and 2013ej, \citealt{Chakraborti14dj,Chakraborti13ej}; \S\ref{SubSec:comparison}), but significantly lower than in \sn{}.
Furthermore, in general $T_e\le T_i$ (e.g., \citealt{Fransson96}, their Figure 5), and in the absence of electron-ion equipartition $T_e\ll T_i$ (\S\ref{SubSec:selfcheksXrays}). 
We thus model the detected X-ray emission from \sn{}\, in the context of a FS origin, and we demonstrate below that this scenario is self consistent. 
%----------------------------------------------------------------
%\subsection{CSM density and evolution of the CSM ionization fraction} 
\subsection{CSM density profile and geometry} 
\label{SubSec:densityXrays}

The broad-band X-ray observations of \sn{}\, provide two complementary ways to estimate the CSM density profile $\rho_{\rm{CSM}}(r)$: first, the normalization of the bremsstrahlung spectrum is directly proportional to the emission measure $EM\equiv \int n_e n_I dV$, where $n_e$ and $n_I$ are the electron and ion number densities in the emitting volume $V$; second, the measured $\rm{NH}_{\rm{int}}$ provides an estimate of the material along the line of sight (Table \ref{Tab:Xcontimuum}). The first method is sensitive to \emph{ionized} material that is responsible for the detected emission, while the second method is sensitive to the column density of \emph{neutral} material ahead of the shock\footnote{The shape of the absorbed continuum of Fig. \ref{Fig:XraySpectraPlot} at $\le$ few keV is consistent with being dominated by a neutral absorber.} and along our line of sight (and does not ``know'' of the three-dimensional geometry of the CSM surrounding the explosion).

We follow the formalism of \cite{Brethauer22}, where a similar model was developed for the strongly interacting SN\,2014C. The \emph{unshocked} CSM density is:
\begin{equation}
\label{Eq:rhoCSMEM}
\rho_{\rm{CSM}}(r)=\frac{m_p}{4} \Big ( \frac{2\times EM(r) \mu_e \mu_I}{V_{\rm{FS}}(r)} \Big )^{1/2} ,
\end{equation}
Where $\mu_e$ and $\mu_I$ are the mean molecular weight per electron and per ion, respectively; $m_p$ is the proton mass; we have assumed a strong shock (for which shock jump conditions dictate a compression factor $\Re=$4 for gas with adiabatic index $\gamma_{ad}=\frac{5}{3}$), and we accounted for the fact that only half of the emitted radiation reaches the observer (following \citealt{Fransson96}).
We parametrize the emitting volume as a portion of a shell of inner radius $R_{in}$ and outer radius $R_{out}$, with $R_{out}\approx 1.2 R_{in}$ and filling factor $f$: $V_{\rm{FS}}=\frac{4\pi}{3}f \big (R_{out}^3-R_{in}^3\big )$. The 1.2 factor is motivated by the self-similar solutions for the FS region for a wind-like CSM (see below) and a stellar progenitor with $\rho_{ej}(r)\propto r^{-n}$ and $n\ge 9$ (for RSGs $n\approx12$), \cite{Chevalier82}. It follows that $ \rho_{\rm{CSM}}\propto f^{-1/2}$. We further assume Solar composition for the CSM (i.e., $\mu_e=1.14$ and $\mu_I=1.24$; \citealt{Zimmerman23} infer a metallicity $Z=0.95\,Z_{\sun}$ for the environment of \sn{}), and an average FS shock velocity $v_{sh}\approx 10^{4}\,\rm{km\,s^{-1}}$. Accounting for the mild shock deceleration during the first $\approx 100$\,d ($v_{sh}\propto t^{-0.1}$ for the self-similar solutions of \citealt{Chevalier82} for $n=12$ and a wind-like CSM) has no impact on the major conclusions from this study.  

The CSM density profile inferred from the emission properties of the bremsstrahlung spectrum is displayed in Fig. \ref{Fig:TemperatureNH}, lower panel, for filling factor $f=1$, and shows a remarkable agreement with a wind-like density profile $\rho_{\rm{CSM}}\propto r^{-2}$ at $r\ge 10^{15}\,\rm{cm}$. At $r<10^{15}\,\rm{cm}$, there is a clear deviation from a wind-density profile. This behavior is mirrored by the measured $\rm{NH_{int}}(t)$: while at late times ($\delta t>10$\,d), $\rm{NH_{int}}(t)$ displays the $\rm{NH_{int}}(t)\propto t^{-1}$ evolution that is expected for a wind-density profile, the very early evolution is significantly steeper ($\rm{NH_{int}}(t)\propto t^{-2.3}$), and suggests a deviation from a wind-density profile with constant level of ionization along the line of sight. 

There are three main ways to reconcile the apparent discrepancy between the temporal evolution of the total H column density calculated from Eq. \ref{Eq:rhoCSMEM} and the measured $\rm{NH_{int}}(t)$ at $\delta t<10$\,d: (i) a time-evolving ionization fraction of the CSM ahead of the FS shock, as a consequence of the ionizing power of the FS radiation; (ii) global CSM asymmetries; (iii) presence of CSM clumps in the innermost CSM $r<10^{15}\,\rm{cm}$ along the line of sight. Each of these three factors is likely to play some role in SN\,2023ixf. Below, we try to isolate the main factor that shapes the observed phenomenology. 

A comparison between the measured and EM-inferred $\rm{NH_{int}}$ of Fig. \ref{Fig:TemperatureNH} would lead to the conclusion that the neutral column fraction evolves from $\approx70$\% at 4\,d to $\approx15$\% at later epochs. While the emergence of a line complex around $6.5$\,keV suggests an increasing level of ionization with time (\S\ref{SubSec:Feline}), we find it unlikely that the drastic change of neutral column fraction between 4-10\,d is only the result of photoionization by the FS expanding in a wind-density medium (see e.g., the numerical simulations by \citealt{Fransson96}). A more likely scenario is one where we are under-estimating the density along the line of sight. This can either result from a global asymmetry of the CSM (i.e., $f<1$), and/or from clumps of material along the line of sight that affect the $\rm{NH_{int}}$ measure, but leave the $EM$ mostly unaffected. The rapid evolution of the optical spectropolarimetric properties of \sn{}\, at early times (\citealt{Vasylyev23}) points at clear deviations from spherical symmetry, while the detection of Fe emission in the X-ray spectra indicates the presence of higher density regions (of a factor $\approx 25$ compared to the X-ray continuum producing medium, \S\ref{SubSec:Feline}). Figure \ref{Fig:TemperatureNH} (lower panel, solid lines), shows two examples of density profiles along the line of sight that would lead to a constant neutral hydrogen column fraction of $\approx15\%$ at all radii, including at $r<10^{15}\,\rm{cm}$. If attributed to a global CSM asymmetry, the implied filling factors would be unphysically large $f\approx 25-400$. Instead, density clumps are more likely, and we note that the overdensity of a factor $\approx 20$ at small radii along the line of sight is intriguingly similar to the inferred density of the Fe-line emitting material.

Finally, the constraints on the column density of ionized CSM material imply an electron scattering optical depth $\tau_{es}\le 0.2$ at $\delta t\ge4$\,d for the ``average'' medium and  $\tau_{es}\le 1$ at $\delta t\ge4$\,d for medium along the line of sight.  
The detection of emission lines with prominent Lorentzian wings in the optical/UV requires instead significantly larger $\tau_{es}> $ a few at $\delta t\lesssim 10$\,d (e.g., \citealt{JacobsonGalan23}), from which we conclude that the region of optical/UV line formation is distinct from the X-ray emitting region even when the two types of emission co-exist in time. This finding translates into density estimates that differ by a factor $\approx 10-100$ at the \emph{same} radii $r\le 10^{15}\,\rm{cm}$ (Fig. \ref{Fig:csm-density-profile}), with the optical/UV emission originating from the densest regions (discussed in \S\ref{SubSec:DiscDensity}). Parenthetically, the low $\tau_{es}$ implies that Comptonization of the most energetic X-ray photons has no dominant effect in the X-ray emitting regions.

To conclude this section, our modeling of the X-ray emission from \sn{}\, thus suggests an anisotropic and/or clumpy X-ray emitting CSM. The overall density profile is not too dissimilar from a wind at $10^{15}<r<10^{16}\,\rm{cm}$. However, the innermost CSM region at  $r<10^{15}\,\rm{cm}$ is considerably more complex with overdensities and deviations from spherical symmetry.
 
 %----------------------------------------------------------------
\subsection{Inferences from the Fe K$\alpha$ line}\label{SubSec:Feline}
At the measured $T_e \approx 5\times 10^{8}$\,K free-free cooling dominates (e.g., \citealt{Fransson96}) and no prominent emission line is expected. However, the X-ray spectra of Fig. \ref{Fig:XraySpectraPlot} show a prominent feature around 6--7\,keV that we interpret as Fe K$\alpha$ emission. For line emission to be prominent, gas temperatures  $T\lesssim 2\times 10^{7}\,\rm{K}$ significantly lower than those implied by the X-ray continuum of \sn{}\, are needed, or, alternatively, a medium with 5--10 times Solar metallicity (which we consider unlikely given the metallicity inference of $0.95\,\rm{Z_{\odot}}$ by  \citealt{Zimmerman23}). A clumpy CSM, with the Fe K$\alpha$ line formed inside the higher-density clumps (i.e., a multi-phase plasma), can explain the observation of the prominent Fe K$\alpha$ line emission \emph{and} a very hot continuum, as has been invoked for other strongly interacting SNe \citep{Kohmura94,Houck98,Uno02,Dwarkadas10,Chandra12,Margutti17,Brethauer22}.
In this scenario, a lower shock temperature can be associated with a slower shock velocity inside clumps. Assuming approximate pressure balance between the clumps and the background medium and for $T\propto v_{\rm{sh}}^{1/2}$, clumps with density $\rho_c \sim (T_{shock}/T_{clump}) \rho_{\rm CSM} \approx 25\rho_{\rm{CSM,sh}}$ (where $\rho_{\rm{CSM,sh}}$ is the density of the medium that produces the X-ray continuum) would satisfy the temperature requirement to produce Fe emission. Interestingly, $\rho_c$ is comparable to the density enhancement required by the $\rm{NH_{int}}$ and density analysis, and it is closer to the density of the optical/UV line forming region (\S\ref{SubSec:csm-density-profile}).

Finally, we comment on the evolution of the Fe K$\alpha$ emission (Table \ref{Tab:FeLine}). The evolution from unresolved neutral Fe K$\alpha$ line to a much broader, resolved spectral feature is consistent with the increasing level of ionization of the CSM  (i.e., Fe atoms at different ionization stages contribute to the detected K$\alpha$ line complex at later times). 
%

%----------------------------------------------------------------
\subsection{Self-consistency checks}\label{SubSec:selfcheksXrays}
In this subsection, we first discuss the expected signature of Inverse Compton (IC) emission on the X-ray spectrum and the nature of the RS and FS in \sn{}\, (i.e. radiative vs. non-radiative) that follow from the properties of the medium inferred above, and we demonstrate that the picture of a FS-dominated thermal X-ray emission is self-consistent. We then address the measured vs. expected FS electron temperatures under different physical scenarios. 

%----------------------------------------------------------------
\subsection{Inverse Compton emission and shock microphysical parameters}\label{SubSec:IC}
Non-thermal X-ray emission in SNe can originate from the upscattering of photospheric optical/UV photons off of relativistic electrons accelerated at the FS through Inverse Compton (IC) processes (e.g., \citealt{Bjornsson04,chevalier2006}). Scattering from a population of non-thermal electrons\footnote{We do not consider here the X-ray emission from the IC scattering by the thermal electron population of electrons as this component tends to be important only for $T_e>10^9$\,K \citep{Fransson96}.} with energies $E=\gamma_e m_e c^2$ (where $\gamma_e$ is the electron Lorentz factor) distributed as $dN/dE \propto E^{-p}$ generates emission with a spectrum $F_{\nu}\propto \nu^{-(p-1)/2}$ in the optically thin regime. The normalization of the resulting IC spectrum further depends on the properties of the seed photons (i.e., photospheric temperature $T_{ph}$ and SN bolometric luminosity $L_{\rm{bol}}$), as well as on the post-shock energy fraction in relativistic electrons $\epsilon_e$, the CSM density $\rho_{\rm{CSM}}$ and shock velocity $v_{sh}$.

We adopt the IC formalism of \cite{Margutti12} modified to (i) accommodate a shock velocity $v_{sh}\propto t^{-\alpha}$ with $\alpha\approx 0.1$ (as expected for a massive stellar progenitor star and a wind-like CSM density profile, \citealt{Chevalier82}) with a normalization chosen to give an average shock velocity of $\sim 10,000\,\rm{km\,s^{-1}}$ in the first $\sim200$\,d of evolution as found from VLBI observations of \sn{}\, (M. Bietenholz, private communication); 
(ii) we also include the effects of IC cooling. 

We use the $L_{\rm{bol}}$ and $T_{ph}$ estimates from  \cite{Zimmerman23}, and we adopt $p=3$ as typically inferred for radio SNe (e.g., \citealt{Soderberg05,chevalier2006,Soderberg10}),  and consistent with the updated predictions of non-linear diffusive shock acceleration (DSA) of particles \citep{Diesing21}. A harder electron distribution with $p\approx 2$ (as expected from classical DSA theory in the test particle limit; e.g., \citealt{Caprioli23} and references therein) would lead to the same conclusions. The luminous $L_{\rm{bol}}$ of \sn{}\, implies that during the first $\approx$ week the electron population is effectively cooling through IC emission (discussed in more detail in \S\ref{SubSec:electron-cooling}). 
After accounting for IC cooling, we find the resulting IC $L_X\ll10^{40}\,\rm{erg\,s^{-1}}$ at all times for all physical choices of $\epsilon_e$ values, and for the CSM densities of interest. This result is consistent with the lack of observational evidence of any non-thermal X-ray component in our spectra. However, it also leaves the parameter space of shock microphysical parameter values entirely unconstrained. In the radio/X-ray SN literature, equipartition with $\epsilon_e\approx 0.1$ is often assumed, but without strong observational support. The analysis of data from some Galactic SN remnants (e.g., \citealt{Sarbadhicary17,Reynolds21} and references therein) favor significantly lower values\footnote{Albeit we caution that Galactic SNR shocks are significantly slower ($v_{sh}\le 0.01\,c$, e.g., \citealt{Zeng19}) and might therefore probe a different regime of shock particle acceleration than young shocks in type-IIP SNe.} $\epsilon_e\lesssim 0.5\times 10^{-3}$, in line with the theoretical expectations from non-linear DSA at non-relativistic shocks \citep{Park15,Gupta24}. In the following, we adopt $\epsilon_e= 10^{-4}$ when needed~\citep[see also][for the similar argument]{murase2014}, and we comment on the implications of this assumption where relevant.
%-------------------------------------
\subsubsection{Radiative vs. Adiabatic Shocks}
\label{SubSubSec:RadAdiabShock}
At $T_e\gtrsim2.6\times 10^7\,$K free-free continuum emission dominates the cooling, and the cooling time-scale derived from \citealt{rybicki1979}, their Eq. 5.15b,  is:
\begin{equation}
\label{Eq:tcoolBremss}
t_{ff}\approx 1.5\times 10^{11} \Big ( \frac{T_e^{1/2}}{n_I \overline{g_{ff}}\mathcal{Z}^2}  \Big) \,\rm{s}
\end{equation}
Where $\overline{g_{ff}}$ is the velocity averaged Gaunt factor and $\mathcal{Z}$ is the charge (effectively, the average number of free electrons per ion in the plasma, i.e., $\approx n_e/n_I=\mu_I/\mu_e$).   For the electron temperatures measured in \sn{}\, and our inferred densities, we find that $t_{ff}\ge t_{\rm{dyn}}$ at all times of our monitoring. We note that IC losses are also not important at $\delta t\ge8$\,d given the optical-UV bolometric luminosity of \sn{}\, from \cite{Zimmerman23}.
We thus conclude that the FS is adiabatic at $\delta t\ge 8$\,d and that a single-temperature spectral modeling is likely adequate  (as opposed to radiative shocks where the spectrum can be formed in gas zones with very different temperatures, \citealt{Nymark06}).\footnote{The same conclusion would be reached by using the $t_{ff}$ relation provided by \cite{Chevalier17}.} 
It follows that at $\delta t\ge 4$\,d the FS X-ray luminosity does not track the shock kinetic luminosity, i.e., the relation $L_X=(4\pi R_{\rm{sh}}^2)\times \frac{1}{2}\rho_{\rm{sh}} v^{3}_{\rm{sh}}$ used by \cite{Zimmerman23} does not apply and the derived inferences are not physical. 

Adopting the shock jump conditions for a strong shock in the thin shell approximation, the density of the shocked ejecta behind the RS is much higher than the density of the shocked CSM behind the forward shock (e.g., \citealt{Fransson96}, their Eq. 3.3): 
\begin{equation}
\rho_{\rm RS}=\frac{(n-3)(n-4)}{(3-s)(4-s)}\times \rho_{\rm FS} \overset{\tiny s=2, n=12}{=} 144\times \rho_{\rm FS}
\end{equation}
The larger density and lower temperature of the RS region (Eq. \ref{Eq:TiRS}) imply significantly smaller free-free cooling times. From Eq. \ref{Eq:tcoolBremss}, we find $t_{ff}\le 0.1 t_{\rm{dyn}}$ for the RS at all times of our monitoring, from which we conclude that the RS is radiative.\footnote{Equation \ref{Eq:tcoolBremss} only includes free-free cooling. Line emission, which becomes important at $T_e\lesssim 2.6\times 10^{7}$\,K, would further increase the cooling rate (e.g., \citealt{Chevalier17}, their Eq. 15), thus reinforcing our conclusions of a radiative RS.} We note that the radiative nature of the RS is independently supported by the detection of the H$\alpha$ line with a boxy profile  \citep{Chevalier1985} in the optical spectra of \sn{} \citep{Teja23}. Radiative cooling of the RS, and the thermal instabilities that may occur as a consequence, lead to the formation of a cold dense shell (CDS) close to the contact discontinuity between the two shocked regions (e.g., \citealt{Chevalier1985, Fransson1984}). The CDS provides a time-dependent photo-electric absorption component that is able to efficiently obscure the entire X-ray emission from the RS, as we show below. 

The column density of the CDS gas is $N_{\rm{CDS}}(t)\approx \frac{M_{\rm{RS}}(t)}{4\pi R(t)_{\rm{sh}}^2 \mu m_p}$, where $M_{\rm{RS}}(t)$ is the ejecta mass that has crossed the RS at a given time $t$, $\mu$ is the mean mass per particle (i.e. ions and free electrons) behind the RS, and $m_p$ is the proton mass (e.g., \citealt{Nymark06}, their Eq. 50). In the self-similar regime of \cite{Chevalier82}, thin shell approximation, $M_{\rm{RS}}=\frac{(n-4)}{(4-s)}M_{\rm{FS}}$, where $M_{\rm{FS}}$ is the mass of the shocked CSM. For $s\approx 2$, $v_{sh}\approx 10^4\,\rm{km\,s^{-1}}$ and the CSM density inferred in \S\ref{SubSec:densityXrays} we find $N_{\rm{CDS}}(t)\approx \frac{10^{25}}{(t/ \rm{d})}\,\rm{cm^{-2}}$ (e.g., \citealt{Chevalier17}, their eq. 17), which implies $N_{\rm{CDS}}\gtrsim10^{23}\,\rm{cm^{-2}}$ at $\delta t<85\,\rm{d}$ and a photo-electric absorption optical depth $\tau_{\rm{phot}}>1$ for photons with energy $E\lesssim 3$\,keV at all times of our monitoring (while observed spectra clearly have emission below 3 keV, Fig. \ref{Fig:XraySpectraPlot}). Finally, adopting the RS luminosity formulation by \cite{Fransson96}, their Eq. 3.10, self-consistently corrected for the photo-electric absorption term calculated above, we predict that the observed RS X-ray luminosity might become comparable to the FS emission at $\delta t\gtrsim 100\,\rm{d}$, potentially leading to a flattening of the X-ray light-curve.  We conclude that the RS X-ray emission is effectively absorbed by the CDS at $\delta t<85\,\rm{d}$, and that the detected emission from \sn{}\, is from the FS material, consistent with our assumptions above. 
%-------------------------------------
\subsubsection{Electron-ion Equipartition}
\label{SubSubSec:eiEquip}
Next, we address the question of the low electron temperatures $T_e$ compared to the expectations from shock dynamics and shock jump conditions of Equation \ref{Eq:TiFS}. The time scale for Coulomb equipartition between electrons and ions in the shocked region is (Eq.12 of \citealt{Chevalier17}):
\begin{equation}
\label{Eq:teiequip}
t_{e-i,eq}\approx2.5\times 10^{7}\Big ( \frac{n_e}{10^7\,\rm{cm^{-3}}}\Big )^{-1} \Big ( \frac{T}{10^9\,\rm{K}}\Big )^{1.5} \rm{s}
\end{equation}
\citep{Spitzer78,Stepney83}. It follows that for the densities of the shocked CSM inferred above\footnote{For a strong shock in a gas with adiabatic index $\gamma_{ad}=5/3$, $\rho_{\rm{sh,CSM}}=4\rho_{\rm{CSM}}$).} $t_{e-i,eq} \approx t_{\rm{dyn}}$ at $\delta t\le 10$\,d, at which time the electrons are marginally in equipartition, with electrons falling more and more out of equilibrium with time as the density drops with radius. Ion-electron collisions are thus ineffective, and we expect $T_e< T_i$ at the FS. We measure $k_bT_e\le 30$\,keV at $\delta t \ge 20$\,d, which is lower than the ion temperature $k_bT_i\approx 100$\,keV from Eq. \ref{Eq:TiFS}, consistent with this expectation. Parenthetically, this result also implies that plasma instabilities are also inefficient in heating the electrons collisionlessly.
%%%%%%%%%%%%%%%%%%%%%%%%%%%%%%%%%%%%%%%%%%%%%%%%%%%%%%%%%%%%%%%%%
\section{Inferences from Radio observations} \label{Sec:RadioInferences}
\subsection{Radio spectral modeling and physical parameters}
We model the radio spectra of \sn{} at each epoch with a smoothed broken power law (BPL) function \citep{Soderberg2006-2003bg} of the form:
\begin{equation}
    F_{\nu} = F_{\rm pk} \left[ \left( \frac{\nu}{\nu_{\rm pk}} \right)^{\alpha_{1}/s} + \left( \frac{\nu}{\nu_{\rm pk}} \right)^{\alpha_{2}/s} \right]^{s}
\end{equation}
Here, $\alpha_{2}$ represents the optically thick spectral index of the synchrotron self-absorbed (SSA) spectrum. We expect $\alpha_{2} =$ 5/2 if the SSA peak ($\nu_{\rm SSA}$) is greater than the characteristic synchrotron frequency ($\nu_{\rm m}$). $\alpha_{1}$ is the optically thin spectral index which is related to the electron power-law index $p$ as $\alpha_{1}=-(p-1)/2$, where the electron energy power-law distribution is $dN(E)/dE=N_{0}E^{-p}$. The smoothing parameter $s$ determines the sharpness of the transition from optically thick to thin phase. $\nu_{\rm pk}$ and $F_{\rm pk}$ represent the frequency and flux density at which the optically thick and thin power-laws of the SSA spectrum intersect.

We fit the BPL model to each of the spectra, treating $\alpha_{2}$, $\nu_{\rm pk}$, and $F_{\rm pk}$ as free parameters, fixing $s = -1$. Since the optically thin phase of the spectra is not well sampled, we impose the optically thin spectral index to be $\alpha_{1}=-1$ ($p=3$). The best-fit parameters are listed in Table \ref{Tab:radio-shock-parameters}, and the best-fit modeled spectra are presented in Figure \ref{Fig:radioSED-fit}.

The spectra are well-represented by a SSA model with optically thick spectral indices varying from $\alpha_{2} =$ 2.4$\pm$0.1 to 2.9$\pm$0.2. These values broadly agree with the theoretical value of 5/2 within error bars. The peak frequencies cascade to lower values at later epochs with a temporal index of $\nu_{\rm pk} \propto$ $t^{-0.78\pm0.03}$. The peak flux densities increases as $F_{\rm pk} \propto t^{ 1.34\pm0.21}$. We note that the NOEMA 84 GHz flux density measurement at $\delta t \approx 38$\,d (purple star symbol in Fig \ref{Fig:radioSED-fit}) is significantly above the near-simultaneous SED at $\delta t \approx 33$ d and even the SED at $\delta t \approx 57$\,d. Also, the 84 GHz flux density at $\delta t \approx 23$\,d is above the modeled SED at $\delta t \approx 33$\,d, which is contradictory to the temporal evolution of SEDs. We address this further in \S \ref{SubSec:radio-limits} and \S \ref{SubSec:secondary-electrons}.

In a standard SSA scenario, for a wind-like CSM, $F_{\rm pk}$ remains constant for $m = 1$ and decreases with time for smaller values of $m$ \citep{chevalier1998}, where $m$ is the shock deceleration parameter defined as $r \propto t^{m}$. The radio spectra of \sn{} evolve as $F_{\rm pk} \propto t^{ 1.34\pm0.21}$ and do not follow this trend. Even though the optically thin evolution is not well sampled, it is clear that the optically thin flux also increases from $\delta t = 33-89$\,d with an average index of $F_{\rm thin} \propto$ $t^{1.1}$, remaining roughly constant between $\delta t = 89-165$\,d. Optically thin synchrotron flux densities scale as $F_{\rm thin} \propto$ $R^{3} N_{0} B^{\frac{p+1}{2}}$, where $N_{0} \propto B^{2}$ if the energy density of relativistic electrons is a constant fraction ($\epsilon_{\rm e}$) of shock energy density ($\rho_{\rm CSM}v^{2}$). In that case, the observed temporal evolution of optically thin flux densities indicates a CSM density profile shallower than that created by a steady stellar wind. We further explore the CSM density profile by detailed modeling of radio SEDs in \S \ref{SubSec:multi-epoch radio modeling}.

The best-fit $\nu_{\rm pk}$ and $F_{\rm pk}$ from the SSA spectrum can constrain shock radii, velocities, and magnetic fields at multiple epochs post-explosion \citep{chevalier1998}. We use equations 13 and 14 from \cite{chevalier1998} and assume $\epsilon_{\rm e} = \epsilon_{\rm B}$ to estimate $R$, $v$, and $B$ at $\delta t =$ 33, 57, 89, and 165\,d and the values are presented in Table \ref{Tab:radio-shock-parameters}. The mean shock velocities are $v \approx$ 1582 $-$ 2716 km\,s$^{-1}$, significantly lower than the velocities derived from optical line measurements ($v \sim$ 8400 km\,s$^{-1}$) at $\delta t \approx$ 2.4$-$14.4\,d \citep{JacobsonGalan23}. 

An SN ejecta of mass $M_{\rm ej} \approx$ 1 $M_{\odot}$ can be decelerated from 8400 km\,s$^{-1}$ to 1582 km\,s$^{-1}$ by a CSM of mass $\approx$ 4.3 $M_{\odot}$. However, a large fraction of ejecta kinetic energy ($\approx$ $\frac{M_{\rm csm}}{M_{\rm ej}+M_{\rm csm}}$ ) will be dissipated in such an interaction which is not observed in the case of \sn{}. Additionally, such a CSM mass would result in a density $\approx 10^{-13}\, \rm g\,cm^{-3}$ at radius $R < 2.8 \times 10^{15}$ cm, resulting very high free-free optical depth ($\tau_{\rm FFA} >> 1$), resulting non-detection at all the observed VLA frequency bands at $\delta t \approx 33$\,d. Thus, the very small shock velocities inferred from the SSA model are not correct. We explore alternative scenarios below.
\startlongtable
\begin{deluxetable}{lcccccccccc}
\tablecaption{Shock parameters of \sn{} estimated from single epoch radio SED modeling using a BPL function.}
\tablehead{
\colhead{Time$^{\rm{a}}$} & \colhead{$\alpha_{2}$} & \colhead{$\nu_{pk}^{\rm{b}}$} & \colhead{$F_{pk}^{\rm{b}}$}    \\
(d)  &  & (GHz) & ($\mu$Jy)  \\
 }
\startdata
33 & 2.8 $\pm$ 0.3 & 17.6 $\pm$ 1.0 & 565 $\pm$ 25 \\
57 & 2.4 $\pm$ 0.1 & 12.3 $\pm$ 0.4 & 998 $\pm$ 16 \\
89 & 2.9 $\pm$ 0.2 & 8.7 $\pm$ 0.2  & 2387 $\pm$ 40 \\
165& 2.4 $\pm$ 0.2 & 5.2 $\pm$ 0.2  & 3998 $\pm$ 107  \\
\enddata
\tablecomments{$^{\rm{a}}$ With respect to date of explosion. \\ $^{\rm{b}}$ $\nu_{\rm pk}$ and $F_{\rm pk}$ are the intersection of the optically thick and thin power laws of synchrotron spectrum. \\ 
\label{Tab:radio-shock-parameters}}
\end{deluxetable}

\subsection{Simultaneous fit to multi-epoch SEDs}
\label{SubSec:multi-epoch radio modeling}
Shock velocities derived from SSA formulation above could be lower limits if free-free absorption (FFA) determines the observed peak of the SED. In this case, the actual SSA peak would be at much lower frequencies. Here, we attempt to simultaneously model the multi-epoch SEDs under the following assumptions: (1) The blast wave is expanding at velocity $\sim$ 10,000 km\,s$^{-1}$ (2) The density of the CSM is $\rho_{\rm CSM}(r) =$ $\rho_{0} (\frac{1}{r^s})$ with electron temperature $T_{\rm e}$ up to the shock radius at $\delta t \approx165$\,d. (3) The CSM outside this radius is wind-like, and the FFA optical depth due to this medium at 1 GHz is $\tau_{\rm out}$. We use the BPL functional form multiplied with $e^{-\tau_{\rm FFA}}$ to implement FFA (i.e., BPL$\times$e$^{-\tau_{\rm FFA}}$). The $\nu_{\rm pk}$ (GHz) and $F_{\rm pk}$ (Jy) are defined as functions of $R$ and $B$ following \cite[eqns ;][]{chevalier1998} for $p = 3$.  
\begin{equation}
\label{Eqn:peak-flux-ssa}
    F_{\rm pk} = \left[ \left( \frac{R B}{5.046 \times 10^{15}} \right) \left( \frac{\epsilon_{\rm e}}{\epsilon_{\rm B}} \right)^{\frac{5}{19}} \left(\frac{D}{\rm Mpc}\right) ^{\frac{-14}{19}} \right]^{\frac{19}{7}} 
\end{equation}
\begin{equation}
\label{Eqn:peak-freq-ssa}
    \nu_{\rm pk} = \left[3.801 \times 10^{17}  \left( \frac{\epsilon_{\rm e}}{\epsilon_{\rm B}} \right) ^{\frac{3}{19}} F_{\rm pk} ^{\frac{11}{19}} \left( \frac{D}{\rm Mpc} \right)^{\frac{22}{19}} \left( \frac{R}{B} \right)^{-1} \right]^{0.5} 
\end{equation}
The FFA optical depth is given by \citep{rybicki1979}:
\begin{equation}
    \label{eqn:FFA-opticaldepth}
    \tau_{\rm FFA} = \int_{R}^{\infty} 0.018 \times T_{\rm e}^{-3/2} Z^2 \nu^{-2} g_{ff} \, n_e \, n_i dr
\end{equation}
We use the Gaunt factor $g_{\rm ff} = 5$, charge $Z =1$, and $n_{e} = n_{i} = \rho_{\rm CSM}/m_{p}$. The free parameters in this physical model are $\epsilon_{\rm e}$, $\epsilon_{\rm B}$, $\rho_{0}$, $s$, $T_{\rm e}$, and $\tau_{\rm out}$. Here, $\rho_{0}$ and $s$ are defined such that $\rho_{\rm CSM}(r)=\rho_{0} (\frac{r}{R_{0}})^{-s}$, where $R_{0}$ is the shock radius at $t =$ 33.45\,d. Initially, we perform Monte-Carlo Markov Chain (MCMC) fit to multi-epoch SEDs, keeping $\epsilon_{\rm e}$, $\epsilon_{\rm B}$, $\rho_{0}$, $s$, $T_{\rm e}$, and $\tau_{\rm out}$ as free parameters using python package \texttt{emcee} \citep{Foreman-emcee2013}. The best-fit values of the parameters are $\epsilon_{\rm e}=0.19$, $\epsilon_{\rm B}=2.04 \times 10^{-6}$, $T_{\rm e}=4.6 \times 10^{5}$ K, $\rho_{0}=4.06 \times 10^{-17}\, \rm g\,cm^{-3}$, $s=1.31$, and $\tau_{\rm out}=14.71$. We note from the corner plots that the parameters ($\epsilon_{\rm e}$, $\epsilon_{\rm B}$) and ($\rho_{0}$, T$_{e}$) are highly correlated based on their covariance. 

\begin{figure} 
 	\centering
 	\includegraphics[width=0.45\textwidth]{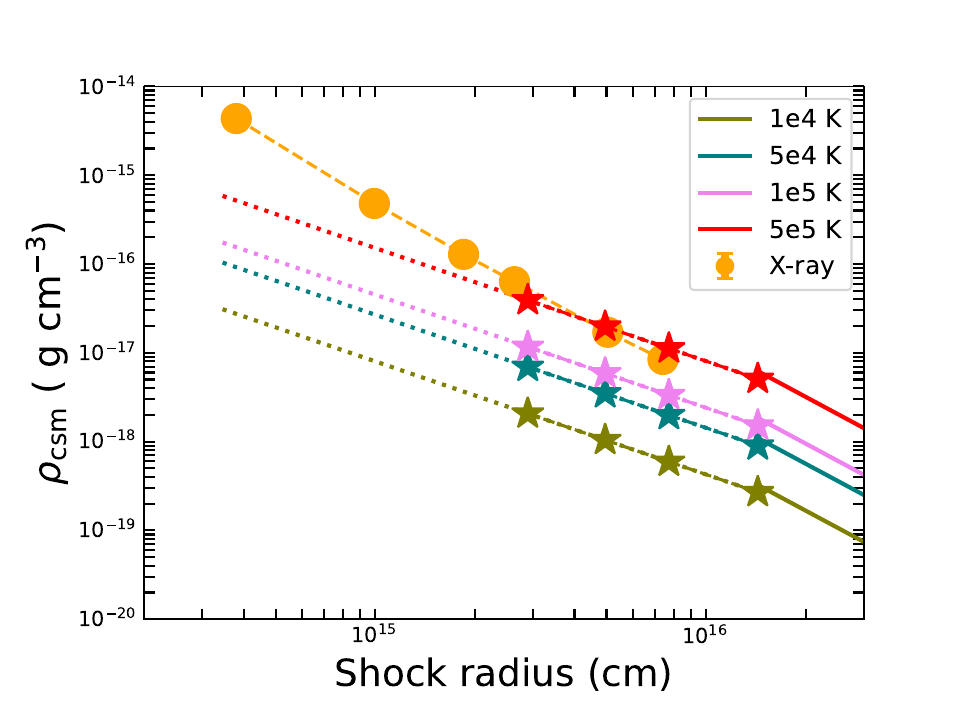}
     \caption{CSM density profiles for various CSM temperatures obtained from multi-epoch radio SED modeling. The star symbols denote the densities at the epoch of radio observations: $t \approx$ 33, 57, 89, and 165\,d, and the solid lines represent the wind density profiles beyond the shock radius at $t \approx$ 165\,d, estimated from the $\tau_{\rm out}$ parameter. The dotted lines are extrapolations of the radio-derived CSM density profiles to earlier epochs.}
 \label{Fig:densities-diff-T}
 \end{figure}

Next, we attempt the fit for different values of $T_{\rm e}=$ 10$^{4}$, 5 $\times$ 10$^{4}$, 10$^{5}$, and 5 $\times$ 10$^{5}$ K to remove the degeneracy between $\rho_{0}$ and T$_{e}$ from the modeling. We allow $\epsilon_{\rm e}$, $\epsilon_{\rm B}$, $\rho_{0}$, $s$, and $\tau_{\rm out}$ to vary. The solutions result in a range of values for $\epsilon_{\rm e} \sim (0.16-0.21)$ and $\epsilon_{\rm B} \sim (0.02-6) \times 10^{-4}$. The best-fit $\rho_{0}$ strongly depend on the value of $T_{\rm e}$, resulting higher $\rho_{0}$ for higher $T_{\rm e}$, while $s$ and $\tau_{\rm out}$ remain roughly similar. We show the variation of best-fit $\rho_{\rm csm}$ for different $T_{\rm e}$ in Figure~\ref{Fig:densities-diff-T}. Interestingly, the densities for $T_{\rm e}=$ 5 $\times$ 10$^{5}$ K model align with the densities derived from the X-ray EM of the previous section. A CSM temperature slightly above $10^{5}$ K is estimated for SN\,1993J at $t >$ 10$-$15\,d post-explosion \citep{Fransson96}, based on modeling the CSM structure using a time-dependent photoionization code \citep{Lundqvist1991}. We thus adopt $T_{\rm e}=$ 5 $\times$ 10$^{5}$ K  as fiducial value.

\begin{figure} 
 	\centering
 	\includegraphics[width=0.5\textwidth]{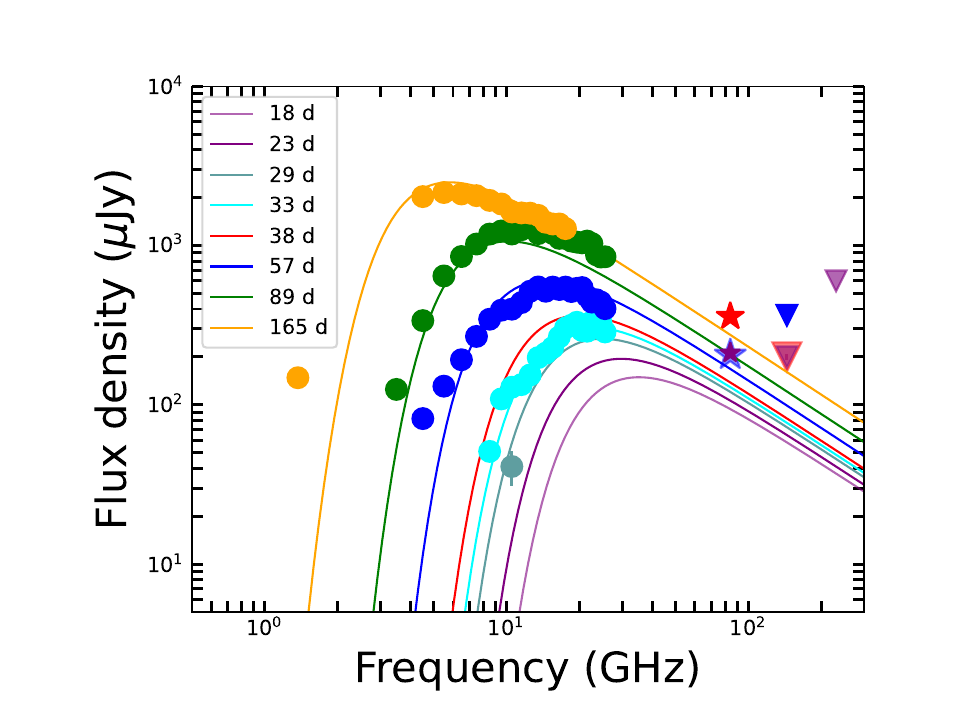}
     \caption{Best-fit radio spectral energy distributions of \sn{} along with observed flux measurements in the time range $\delta t=33-165$\,d. Inverted triangles mark the 5$\sigma$ flux density upper limits. The parameters are obtained by fitting the model, fixing $T_{\rm e}=5 \times 10^{5}$ K and $\epsilon_{\rm e}=10^{-4}$, keeping $\epsilon_{\rm B}$, $\rho_{0}$, $s$, and $\tau_{\rm out}$ as free parameters.}
 \label{Fig:radio-2dfit}
 \end{figure}

\begin{figure} 
 	\centering
  \includegraphics[width=0.5\textwidth]{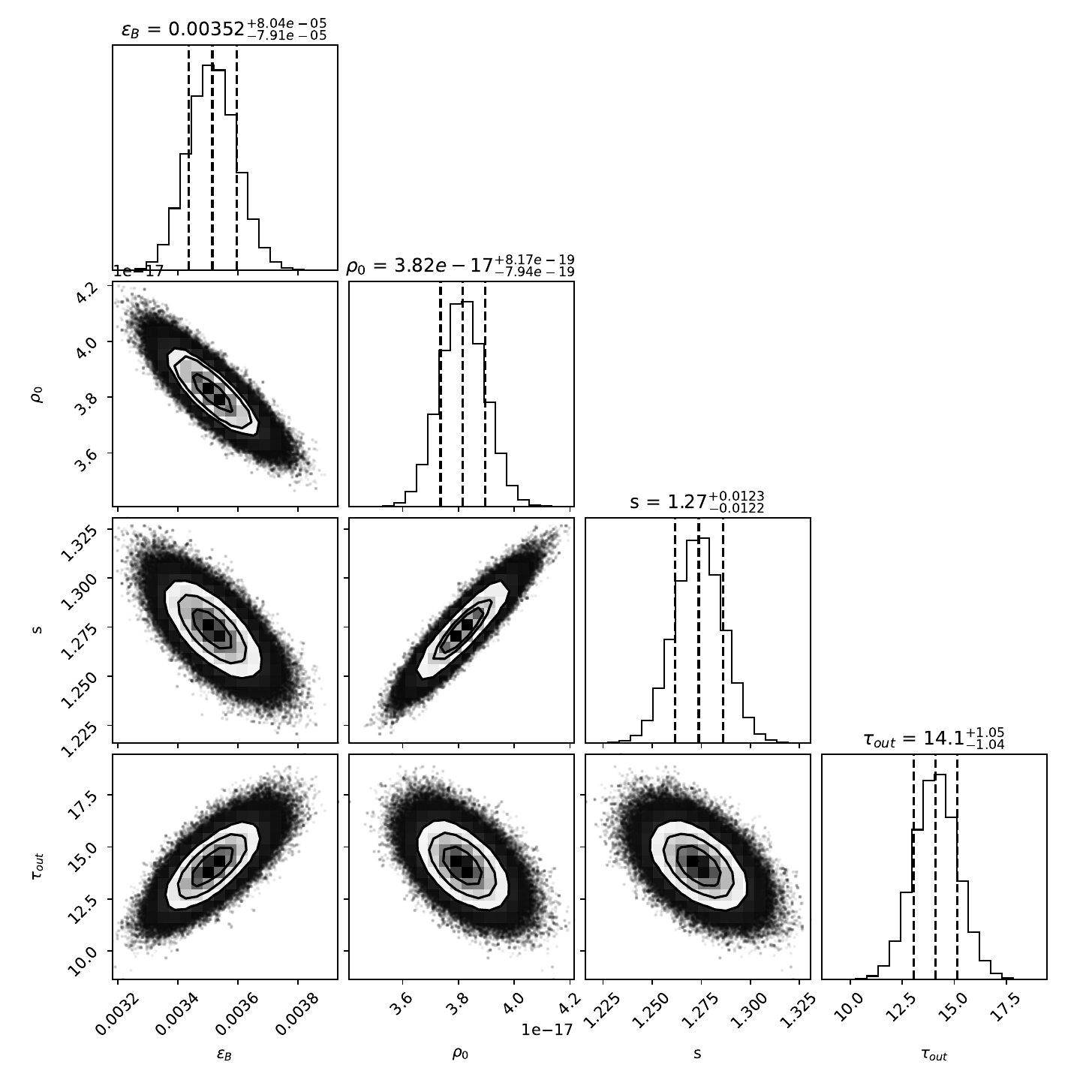}
     \caption{Corner plot showing the results of our MCMC fitting of multi-epoch radio SEDs (Figure \ref{Fig:radio-2dfit}) of \sn{} (see \S \ref{SubSec:multi-epoch radio modeling} for details). $\epsilon_{\rm B}$ is the fraction of shock energy fed into the magnetic fields. $\rho_{0}$ and $s$ are defined such that $\rho_{\rm CSM}(r)=\rho_{0} (\frac{r}{R_{0}})^{-s}$, where $R_{0}$ is the shock radius at $\delta t =$ 33.45\,d. $\tau_{\rm out}$ is the FFA optical depth at 1 GHz due to the medium outside the shock radius at $\delta t = 165.78$\,d. The corner plot is obtained by fitting the model, fixing $T_{\rm e}=5 \times 10^{5}$ K and $\epsilon_{\rm e}=10^{-4}$ and keeping $\epsilon_{\rm B}$, $\rho_{0}$, $s$, and $\tau_{\rm out}$ as free parameters. In each panel, the 16, 50, and 84 percentiles are marked.}
 \label{Fig:radio-2dfit-corner-plot}
 \end{figure}

Finally, we fit the model for different values of $\epsilon_{\rm e}$ fixing $T_{\rm e} =$ 5 $\times$ 10$^{5}$ K, and allowing $\epsilon_{\rm B}$, $\rho_{0}$, $s$, and $\tau_{\rm out}$ vary. The best-fit solutions are presented in Table \ref{Tab:multi-epoch-sed-para}, with the corresponding spectra shown in Figure \ref{Fig:radio-2dfit}, and the corner plot in Figure \ref{Fig:radio-2dfit-corner-plot}. The results show a locus of $\epsilon_{\rm e}$ and $\epsilon_{\rm B}$ values that can account for the observed SEDs for a similar density profile. For further analysis and interpretation, we select the solution with $\epsilon_{\rm e} = 10^{-4}$, based on the argument of \S \ref{SubSec:IC}. We note that the observed data show optically thick spectral slopes that are flatter than those predicted by the best-fit FFA model. This is likely due to inhomogeneities that soften the optically thick portion of the spectrum, whereas the FFA model assumes a uniform optical depth.

\startlongtable
\begin{deluxetable*}{ccccccccccc}
\tablecaption{Best-fit parameters estimated from multi-epoch radio SED modeling discussed in \S \ref{SubSec:multi-epoch radio modeling}.}
\tablehead{
\colhead{T$_e$} & \colhead{$\epsilon_{e}$} & \colhead{$\epsilon_{B}$} & \colhead{$\rho_{\rm 0}$} & $s$ & $\tau_{\rm out}$ & $B$ ($\delta t$ = 33, 57, 89, 165 d)  \\
(K)  & - & - & ($\times$10$^{-17}$gm\,cm$^{-3}$) & - & (at 1 GHz) & (Gauss)
 }
\startdata
5 $\times$ 10$^{5}$ & 0.1 & 3.42$^{+0.08}_{-0.07}$ $\times$ 10$^{-6}$ & 3.91$^{+0.08}_{-0.08}$ & 1.27$^{0.01}_{-0.01}$ & 16.70$^{+1.07}_{-1.05}$ & 0.058, 0.041, 0.031, 0.021 \\
5 $\times$ 10$^{5}$ & 0.01 & 3.43$^{+0.08}_{-0.07}$ $\times$ 10$^{-5}$  &  3.89$^{+0.08}_{-0.08}$ & 1.27$^{0.01}_{-0.01}$ & 16.20$^{+1.05}_{-1.04}$ & 0.183, 0.130, 0.098, 0.066 \\
5 $\times$ 10$^{5}$ & 0.001 & 3.46$^{+0.08}_{-0.07}$ $\times$ 10$^{-4}$ &  3.86$^{+0.08}_{-0.08}$ &  1.27$^{0.01}_{-0.01}$ & 15.40$^{+1.05}_{-1.04}$ & 0.579, 0.412, 0.311, 0.210 \\
5 $\times$ 10$^{5}$ & 0.0001 & 3.52$^{+0.08}_{-0.08}$ $\times$ 10$^{-3}$ & 3.82$^{+0.08}_{-0.08}$ & 1.27$^{0.01}_{-0.01}$ & 14.10$^{+1.05}_{-1.04}$ & 1.838, 1.306, 0.985, 0.665 \\
5 $\times$ 10$^{5}$ & 10$^{-5}$ & 3.61$^{+0.09}_{-0.08}$ $\times$ 10$^{-2}$ & 3.74$^{+0.08}_{-0.08}$ & 1.27$^{0.01}_{-0.01}$ & 12.00$^{+1.05}_{-1.04}$ & 5.825, 4.138, 3.123, 2.108 \\
\enddata
\tablecomments{$T_{\rm e}$ is the CSM electron temperature. $\epsilon_{\rm e}$ and $\epsilon_{\rm B}$ are the fractions of shock energy fed into relativistic electrons and magnetic fields, respectively. $\rho_{0}$ and $s$ are defined such that $\rho_{\rm CSM}(r)=\rho_{0} (\frac{r}{R_{0}})^{-s}$, where $R_{0}$ is the shock radius at $\delta t =$ 33.45\,d. 
%\km{What is the value of $R_0$? This looks different from $10^{15}$~cm.}
$\tau_{\rm out}$ is the FFA optical depth at 1 GHz due to the medium outside the shock radius at $\delta t = 165.78$\,d. 
%\km{I noticed that $t$ and $\delta t$ are used in a mixed manner. Please fix the notations.}
The listed parameters are obtained by fitting the model, fixing $T_{\rm e}$ and $\epsilon_{\rm e}$, keeping $\epsilon_{\rm B}$, $\rho_{0}$, $s$, and $\tau_{\rm out}$ as free parameters. $B$ is the magnetic field estimated using the scaling relation $\frac{B^{2}}{8\pi}=\epsilon_{\rm B}\rho_{\rm CSM}(r) v^{2}$. \\ 
\label{Tab:multi-epoch-sed-para}}
\end{deluxetable*}
 
\subsection{Energy losses of non-thermal electrons}
\label{SubSec:electron-cooling}
Synchrotron-emitting electrons can lose energy through various processes: synchrotron cooling, Inverse Compton (IC) cooling, and Coulomb cooling. These processes become significant when the energy loss timescales are comparable to or shorter than the dynamical timescale ($t_{\rm d}$) of the system. The timescales for synchrotron ($t_{\rm sync}$), IC ($t_{\rm IC}$), and Coulomb cooling ($t_{\rm Coul}$) are given by \citep{Chevalier17,Fransson96}:
\begin{equation}
    t_{\rm sync} = 7.74 \times 10^{8} \, \gamma^{-1} B^{-2} \, \rm sec
\end{equation}
 \begin{equation}
    t_{\rm IC} = 1.16 \times 10^{19} \, \gamma^{-1} L_{\rm bol}^{-1} \, R^{2} \, \rm sec
\end{equation}

\begin{equation}
    t_{\rm Coul} = 0.67 \times 10^{-12} \, \gamma \rho_{\rm csm}^{-1} \, \rm sec
\end{equation}
$L_{\rm bol}$ is the bolometric luminosity of the SN, which includes contributions from both the ejecta and circumstellar interaction. If the intrinsic spectrum of relativistic electrons is dN(E)/dE $= N_{0} E^{-p}$, the synchrotron and IC cooling will steepen the distribution of these electrons to $N_{0} E^{-p-1}$. Consequently, the synchrotron spectrum will have a steeper spectral index ($F_{\nu} \propto \nu^{-p/2}$) above a certain frequency, known as the synchrotron/IC cooling frequency ($\nu_{\rm sync}$/$\nu_{\rm IC}$). This is the characteristic synchrotron emission frequency ($\nu = 1.2 \times 10^{6} B \gamma^{2}$ Hz) of an electron with Lorentz factor ($\gamma$) for which $t_{SC}$ or $t_{IC}$ equals the dynamical timescale. On the other hand, Coulomb losses alter the spectrum in the opposite direction, resulting in a flatter spectral index ($F_{\nu} \propto \nu^{-p/2 + 1}$) at frequencies below a characteristic cooling frequency ($\nu_{\rm Coul}$) where $t_{\rm Coul} = t_{\rm d}$. We caution that the steepening/flattening of the spectrum can be a bit more complicated since it depends on multiple populations of electrons accelerated at different times \citep{Diesing2019} and the calculation that we follow here is a simplified treatment.
 
Depending on the values of $\epsilon_{\rm e}$, $\epsilon_{\rm B}$, and the CSM density profile, the cooling frequencies and their evolution across the spectrum will vary. We calculate these cooling frequencies to understand the possible impact of electron cooling on the modeled SEDs and best-fit parameters. 
We use the magnetic field values from the scaling relation $\frac{B^{2}}{8\pi} = \epsilon_{\rm B}\,\rho\,v^{2}$ to estimate the $\nu_{\rm sync}$. %\footnote{If one uses the downstream energy density we should use $B^2=9\pi \rho v^2$}. 
To calculate $\nu_{\rm IC}$, we use the bolometric luminosities from \cite{Zimmerman23}. Various cooling frequencies for the model with $\epsilon_{\rm e} = 10^{-4}$ are presented in Table \ref{Tab:cooling-frequencies}. 
\begin{deluxetable}{cccccc}
\tablecaption{ Cooling frequencies that correspond to the best-fit parameters from multi-epoch radio SED modeling listed in Table \ref{Tab:multi-epoch-sed-para} for $\epsilon_{\rm e} = 10^{-4}$.}
\tablehead{
\colhead{Time} &\colhead{$\nu_{\rm sync}$} & \colhead{$\nu_{\rm IC}$} & \colhead{$\nu_{\rm Coul}$} & \colhead{$\nu_{\rm SSA}$} & \colhead{$\nu_{\rm peak}$}\\
 (d) &(GHz)  & (GHz) & (GHz) & (GHz) & (GHz) } 
\startdata
33.45 & 14 & 0.3 & 65 & 6 & 22 \\
57.31 & 13 & 1 & 34 & 4 & 15 \\
89.30 & 13 & 28 & 20 & 4 & 10 \\
165.78 & 12 & 248 & 10 & 2 & 6 \\
\enddata
\tablecomments{$\nu_{\rm sync}$, $\nu_{\rm IC}$, and $\nu_{\rm Coul}$ are cooling frequencies due to synchrotron, IC, and Coulomb losses. $\nu_{\rm SSA}$ is the synchrotron self-absorption frequency. At $\nu > \nu_{\rm IC}$, IC cooling is important and at $\nu > \nu_{\rm sync}$, synchrotron cooling is important whereas Coulomb cooling becomes important at $\nu < \nu_{\rm Coul}$. $\nu_{\rm peak}$ is the observed peak of the SED defined by free-free absorption.
\label{Tab:cooling-frequencies}}
\end{deluxetable}

Except for $t \approx$ 165\,d, the competing effects of synchrotron/IC cooling and Coulomb cooling could be affecting the observed spectrum, resulting in no clear signature of either. At $t \approx$ 165\,d, a spectral steepening due to synchrotron cooling is expected at $\nu =$ 12.2 GHz. Although our observations do not adequately cover the optically thin phase of evolution, there is no evident spectral steepening in any of the spectra. We compute the cooling frequencies for the model with $\epsilon_{\rm e} = 10^{-5}$ and find that there is synchrotron and/or IC cooling at frequencies $>$ 0.4 GHz and Coulomb cooling at frequencies $<$ 29 GHz. However, as mentioned previously, these effects could cancel each other out.

To summarize, cooling processes can modify the intrinsic SSA from what is defined by equations \ref{Eqn:peak-flux-ssa} and \ref{Eqn:peak-freq-ssa}. Ideally, one needs to systematically incorporate all the cooling losses in the electron distribution and model the radio emission \citep{Fransson1998}. However, since FFA dominates the absorption, our density estimates from radio modeling are not significantly impacted by the shape of intrinsic SSA. We quantitatively address this in \S \ref{SubSec: Densities-model agnostic approach}.
\subsection{CSM densities from FFA model}
\label{SubSec: Densities-model agnostic approach}
The density of the CSM can be estimated by fitting an FFA-only model to each of the spectra. In this approach, the derived densities are agnostic about the shape of the intrinsic synchrotron spectrum. We model the single epoch SEDs with the FFA model of the form:
\begin{equation}
    F_{\nu} = K_{1} \left( \frac{\nu}{5\, \rm GHz} \right)^{\beta} \, \rm exp \left[ -K_{2} \left( \frac{\nu}{5\, \rm GHz} \right)^{-2.1} \right]
\end{equation}
Here $K_{1}$ and $K_{2}$ are flux density and optical depth normalization constants, respectively. $\beta$ is the optically thin spectral index. We fix $\beta =-1$ and fit the function, allowing $K_{1}$ and $K_{2}$ to vary as free parameters for each spectrum. 

\begin{figure} 
 	\centering
  \includegraphics[width=0.5\textwidth]{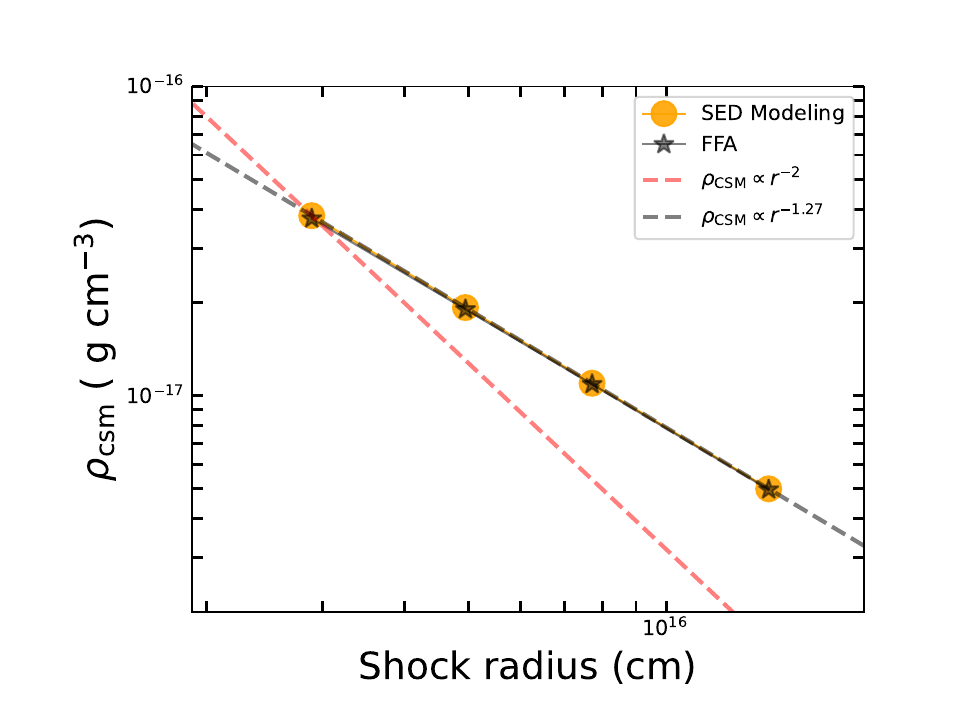}
     \caption{Best-fit CSM densities of \sn{} derived from multi-epoch radio SED modeling (orange circles) and from single-epoch FFA only fits (black star symbols). Dashed red and black lines are to guide the eye and denote a reference wind-density profile $\rho_{\rm CSM}(r) \propto r^{-2}$ and $\rho_{\rm CSM}(r) \propto r^{-1.27}$, respectively.}
 \label{Fig:density-FFAonly-radioSED}
 \end{figure}

We use the $\tau_{\rm FFA}$ (5 GHz) to fit for the CSM density profile following equation \ref{eqn:FFA-opticaldepth} assuming $T_{\rm e} = 5 \times 10^{5}$ K. The best-fit density profile is $\rho_{\rm CSM}(r)= \rho_{0} (\frac{r}{R_{0}})^{-1.26}$, where $\rho_{0} =$ 3.74 $\times$ 10$^{-17} \rm g\,cm^{-3}$ for the shock radius at $\delta t=33.45$\,d. This density is consistent with the densities derived from multi-epoch radio modeling for the range of $\epsilon_{\rm e} =$ 0.1 $-$ 10$^{-5}$. Fig \ref{Fig:density-FFAonly-radioSED} shows a comparison between CSM densities derived by modeling multi-epoch radio SEDs for $\epsilon_{\rm e} = 10^{-4}$ and that from single epoch FFA fits. 
 
\subsection{Early radio upper limits}
\label{SubSec:radio-limits}
\begin{figure*} 
 	\centering
  \includegraphics[width=0.49\textwidth]{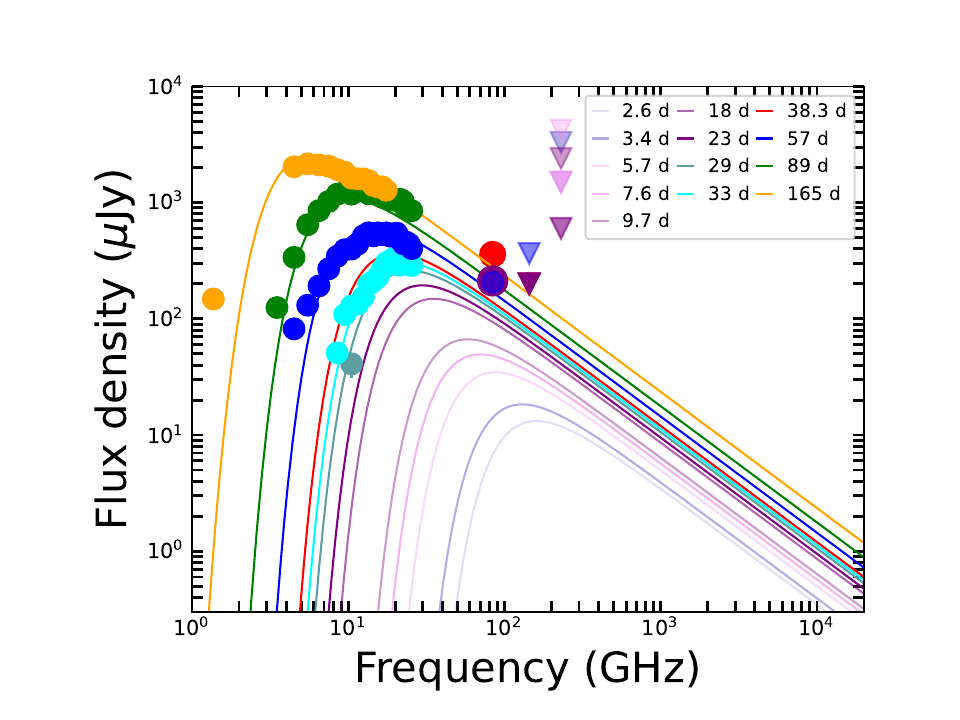}
   \includegraphics[width=0.49\textwidth]{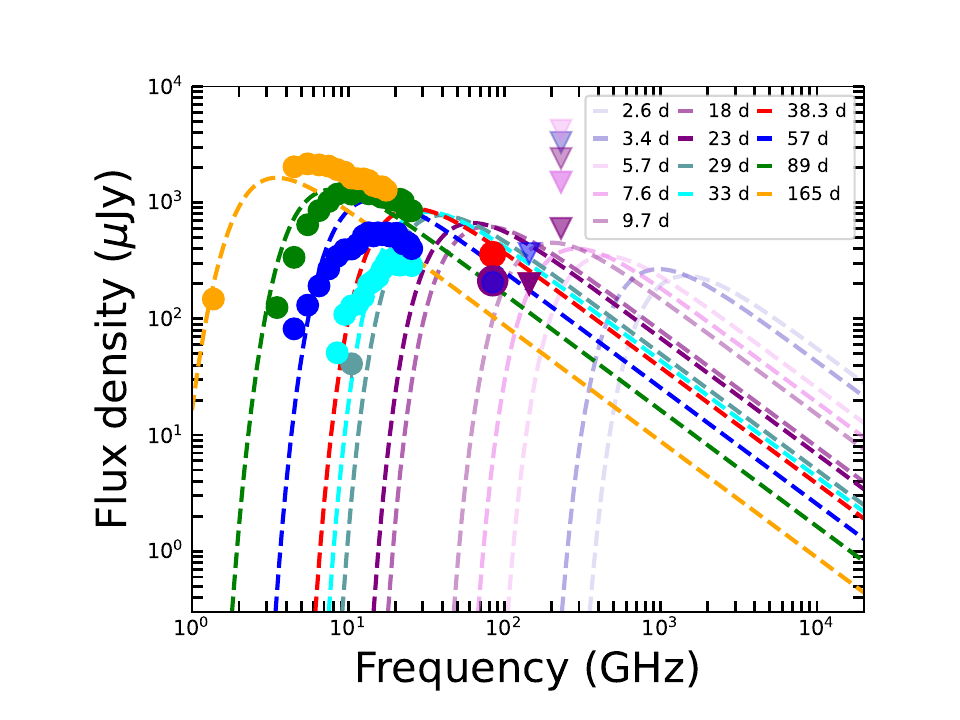}
     \caption{Left panel: Best-fit radio SEDs from the multi-epoch spectral analysis presented in \S \ref{SubSec:multi-epoch radio modeling} for a CSM density profile $\rho_{\rm CSM}(r) \propto r^{-1.27}$ extrapolated to earlier times. Right panel: Expected radio SEDs that could arise from X-ray emitting plasma for a CSM density profile $\rho_{\rm CSM}(r) \propto r^{-2}$.}
 \label{Fig:radio-limits-model}
 \end{figure*}

Multi-epoch ($\delta t \approx 33-165$\,d) radio SED modeling favors a CSM density profile $\rho_{\rm CSM}(r) \propto r^{-1.27}$ whereas the densities derived from X-ray observations ($\delta t \approx 4-86$ d) follow a density profile $\rho_{\rm CSM}(r) \propto r^{-2}$. Either both density regions co-exist in the regions probed by these observations, or there is a transition from $\rho_{\rm CSM} \propto r^{-2}$ to $\rho_{\rm CSM} \propto r^{-1.27}$ from smaller to larger shock radii. To check this, we simulate the radio SEDs that could arise from the X-ray emitting plasma for $\epsilon_{\rm e} = 10^{-4}$ and $\epsilon_{\rm B} = 3.52 \times 10^{-3}$ (see Fig. \ref{Fig:radio-limits-model}). We also extrapolate the best-fit radio model in time to earlier epochs covering the X-ray observation epochs. Additionally, we place the early time ($\delta t \approx 2-18$\,d) radio flux density upper limits from the literature \citep{Berger23} on these two sets of SEDs. At $\delta t \approx 2-18$\,d, the radio upper limits are consistent with both sets of SEDs. At $\delta t \approx 23$\,d, the SED from X-ray emitting plasma is above the 84 GHz NOEMA flux density by a factor of $\approx$ 2. At times $\delta t > 23$\,d, the SEDs from X-ray emitting plasma overpredict the observed radio flux densities in the optically thick regime (see cyan curve for $\delta t \approx 33$\,d in Fig \ref{Fig:radio-limits-model}). The model flux at $\delta t \approx$ 38 and 62\,d are roughly consistent with the predictions from the radio SEDs from X-ray emitting plasma (red and blue circles at 84 GHz). It is likely that $\rho_{\rm CSM} (r) \propto r^{-1.27}$ profile is dominant at radii corresponding to $\delta t \gtrsim 23$ d. However, the extrapolated radio SEDs at $\delta t \approx 23$ and 38\,d underpredict the 84 GHz NOEMA fluxes. 
We explore the possibility of contribution from secondary electrons and positrons to address this in \S \ref{SubSec:secondary-electrons}. 

The flatter density profile from radio SED modeling can be interpreted in two ways. One, the corresponding stellar evolution phase has either varying $\dot{M}$ or $v_{\rm w}$. Two, the CSM is clumpy and non-spherically symmetric. If the clumpy CSM has a filling factor $f_{\rm cl}$, the free-free optical depth will be $\tau_{\rm FFA} \propto \int f_{\rm cl} n_{\rm e} n_{\rm i} dr$ \citep{Fransson96}. If the $f_{\rm cl}$ varies with radius ($f_{\rm cl} \propto r^{1.49}$), a wind density CSM profile can result in the observed radio SEDs.

\subsection{Radio emission from secondary leptons produced by cosmic-ray protons}
\label{SubSec:secondary-electrons}
Efficient cosmic-ray (CR) proton acceleration can happen when the SN ejecta interacts with dense CSM and the shock becomes radiation unmediated~\citep{Murase2011,Katz2012,Murase2018}. The accelerated protons undergo $pp$ collisions and produce gamma rays, neutrinos, and secondary electron-positron pairs. These secondary leptons can produce synchrotron emission detectable at high radio frequencies \citep{murase2014,Petropoulou2016,Murase2024}. In SN\,2023ixf, we note that the 84 GHz flux densities at $t \approx$ 23 and 38\,d are well above the modeled SEDs (see Fig \ref{Fig:radio-2dfit}). Given that the energy fraction carried by cosmic-ray protons, $\epsilon_p\sim0.1$, is larger than that of electrons, $\epsilon_e\sim{10}^{-4}$, the secondary emission could be more prominent when $\epsilon_e \lesssim0.2 \epsilon_{p,-1}{\rm min}[1,f_{pp}]$, where $f_{pp}\simeq2.7\times{10}^{-2}~\rho_{\rm CSM,-16.5}R_{15.5}v_{\rm sh,9}^{-1}$ is the efficiency of inelastic $pp$ interactions. 
Indeed, the NOEMA flux at $t \approx$ 23.0 and 38.2~d are higher than the expectation of the best-fit radio SED model that can be compatible with primary electron acceleration. Thus, we explore the possibility of a contribution from the secondary electron-positron population towards the synchrotron emission at these epochs, adopting the formulation from \cite{murase2014}.

The magnetic field strength is estimated to be $B\simeq1.6~{\rm G}~\epsilon_{B,-2.5}^{1/2}\rho_{\rm CSM,-16.5}^{1/2}v_{\rm sh,9}$ (see also Table~4). 
Noting that the secondary pairs are injected at $\gamma_h\sim68$ instead of $\gamma_{\rm inj}$ for primary electrons, the characteristic frequency of hadronically injected electrons is given by
\begin{equation}
    \nu_{h}\approx \frac{3}{4\pi}\gamma_h^2\frac{eB}{m_e c}\simeq 30~{\rm GHz}~\epsilon_{B,-2.5}^{1/2} \rho_{\rm CSM,-16.5}^{1/2}v_{\rm sh,9},
\end{equation}
and the critical frequency determined by radiative (synchrotorn or IC) cooling is 
\begin{equation}
\nu_{\rm rc}\simeq 0.76~{\rm GHz}~{[(1+Y)/9]}^{-2}\epsilon_{B,-2.5}^{-3/2} \rho_{\rm CSM,-16.5}^{-3/2}R_{15.5}^{-2}v_{\rm sh,9}^{-1}, 
\end{equation}
where $Y$ is introduced as 
\begin{equation}
Y\approx \frac{2L_{\rm bol}}{R^2cB^2}\simeq8.2~L_{\rm bol,42.5}\epsilon_{B,-2.5}^{-1}\rho_{\rm CSM,-16.5}^{-1}R_{15.5}^{-2}v_{\rm sh,9}^{-2}
\end{equation}
in the limit that the system is optically thin to Thomson scattering. 
We estimate $\nu_{h}$ at $\delta t \approx$ 23.0, 38.2, and 62.04 d assuming a shock velocity of 10000 km\,s$^{-1}$ and $\epsilon_{\rm B}= 3.52\times10^{-3}$. 
%We use the CSM densities estimated from X-ray analysis interpolated to these epochs using $\rho_{\rm CSM}(r) \propto r^{-2}$ scaling. 
Then the characteristic frequency of secondary electrons and positrons are $\nu_{h} \sim$ 45, 33, and 24 GHz at $\delta t \approx$ 23, 38, and 62 d.
Note that $\nu_{\rm rc}\approx{\rm min}[\nu_{\rm sync},\nu_{\rm IC}]$ (cf. Table~5). Thanks to the IC cooling, we find $\nu_{\rm rc}\ll \nu_{h}$, i.e., the system at early times is in the fast cooling regime~\citep{Murase2024}, in which the spectrum of electrons and positrons should be extended to energies below $\gamma_h$. 

\citet{Murase2024} presented calculations of radio emission from secondary electrons and positrons, including Coulomb losses and free-free absorption, and showed that the predicted radio fluxes of SN\,2023ixf at $\delta t \approx 20-30$ d can be $F_{\nu} \sim 10-100\, \mu$Jy for $\dot{M} \sim {\rm a~few}\times 10^{-4}\,M_{\odot}\,\rm yr^{-1}$. Note that their shock luminosity is too conservative partly because the self-similar solution is assumed. Our values of $\rho_{\rm CSM}$ and $v_{\rm sh}$ are larger by a factor of 3 and by a factor of 2, respectively. While they assumed that the CSM extends only up to $R_{\rm CSM}=10^{15}$~cm, as shown in this work, the CSM should be extended beyond $10^{15}$~cm. Thus, the predicted radio emission should last longer. For $p>2$, the secondary synchrotron flux in the high-frequency radio band is estimated to be~\citep[e.g., Eq.~65 of][]{Murase2024}
\begin{eqnarray}
F_\nu^{\rm sync} &\approx& \frac{(p-2)\epsilon_pL_s}{48\pi(1+Y)d^2}{\rm min}[1,f_{pp}]g_{\rm Coul}{\left(\frac{\nu}{\nu_{b}}\right)}^{2-\beta_{\rm sync}}\frac{1}{\nu}\nonumber\\
&\simeq& 350~\mu {\rm Jy}~(p-2)\epsilon_{p,-1.3}{[(1+Y)/9]}^{-1}g_{\rm Coul}\nonumber\\
&\times&\rho_{\rm CSM,-16.5}^2R_{15.5}^3 v_{\rm sh,9}^{-2} {(\nu/\nu_b)}^{2-\beta_{\rm sync}}{(100~\rm GHz/\nu)}, \,\,\,\,\,\,\,\,\,\,\,\,
\end{eqnarray}
where $L_s$ is the shock kinetic luminosity of the FS and $f_{pp}<1$ is assumed. Note that $1-\beta_{\rm syn}=-p/2$ for $\nu>\nu_b$ and $\nu_b\approx{\rm max}[\nu_h,\nu_{\rm Coul}]$, and $g_{\rm Coul}$ is the possible correction due to Coulomb losses~\citep{Murase2024}. 
Thus, intriguingly, the flux density predicted in the hadronic scenario for $\epsilon_p\sim0.03-0.1$ is consistent with the observed NOEMA fluxes at $\delta t \approx$ 23.0 and 38.2~d. We note that the synchrotron spectra of the primary electron emission and secondary electron-positron emission are similar at $\nu>\nu_h(>\nu_{\rm inj})$, where the secondary scenario is effectively compatible with the results of the primary scenario with $\epsilon_e \sim \frac{1}{6}\epsilon_p{\rm min}[1,f_{pp}](\gamma_h/\gamma_{\rm inj})^{p-1}$. 
Also, as indicated in Table~5, Coulomb losses are relevant, especially if the absorption is caused by FFA. For $p=3$ and $\nu<\nu_{\rm Coul}$, the secondary scenario predicts $F_\nu\propto \nu^{0}$ while the primary scenario does $F_\nu\propto \nu^{-0.5}$. Apparently, the segment of $F_\nu\propto \nu^0$ seen at $\delta t \approx$ 33 and 57~d is consistent with both primary and secondary scenarios, and more dedicated modeling is necessary to reveal the contribution of proton acceleration in early radio light curves. 
%%%%%%%%%%%%%%%%%%%%%%%%%%%%%%%%%%%%%%%%%%%%%%%%%%%%%%%%%%%%%%%%%%%%%%%%
\section{Discussion} \label{Sec:Disc}
 %----------------------------------------------------------
\subsection{SN\,2023ixf in the context of other radio/X-ray core-collapse supernovae}
\label{SubSec:comparison}
\begin{figure} 
 	\centering
 	\includegraphics[width=0.45\textwidth]{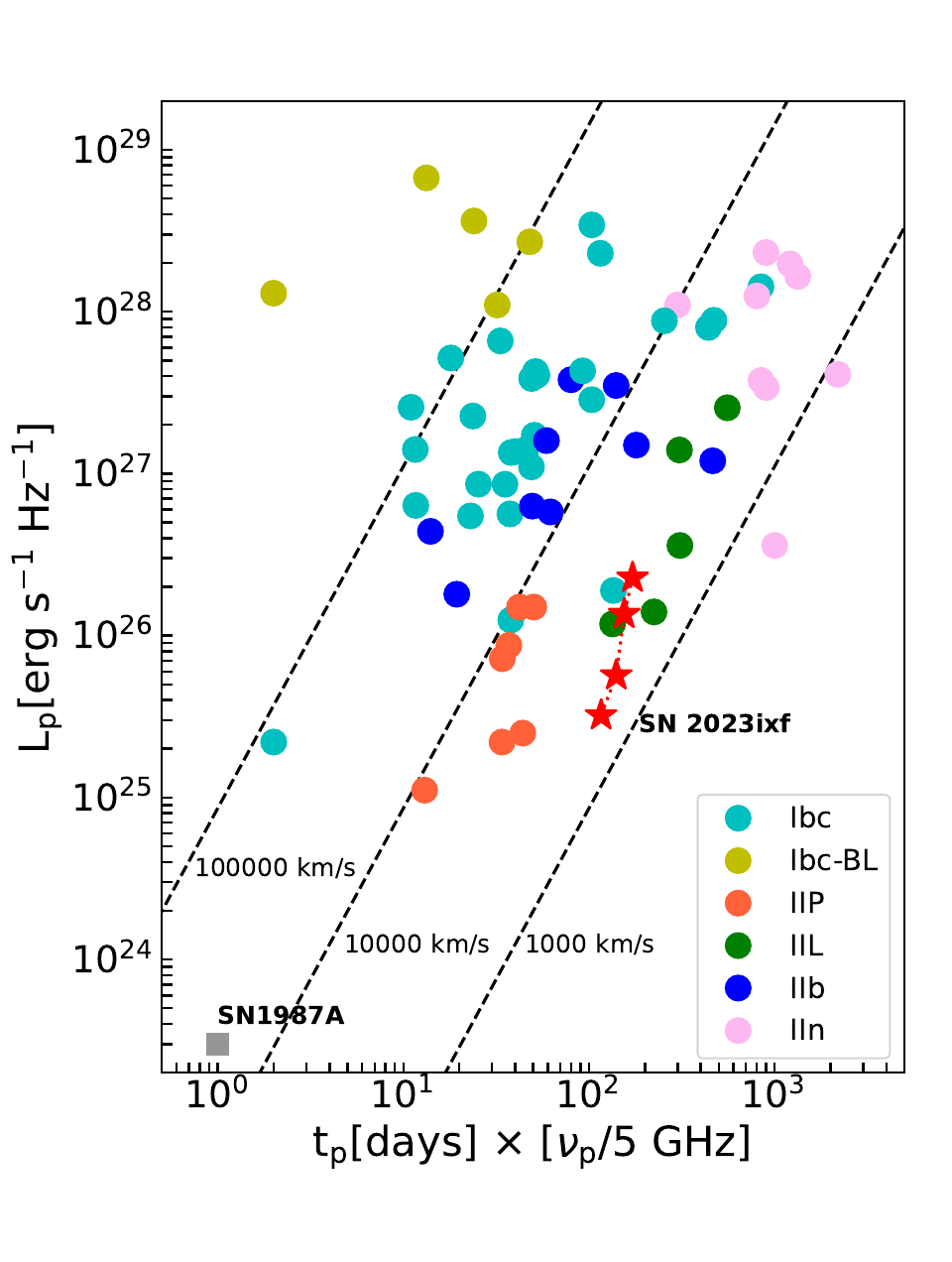} 
  \caption{Peak radio spectral luminosity of \sn{} (red star symbol) in context of other core-collapse supernovae. SNe\, Ibc (cyan circles), SNe\,Ibc-BL (yellow circles), SNe\,IIP (red circles), SNe\,IIb (blue circle), and SNe\,IIn (pink circles). Data are collected from \citep{Bietenholz2021} and references therein. The dashed line denotes the mean shock velocity in a standard synchrotron self-absorption (SSA) scenario \citep{chevalier1998}. }
 \label{Fig:radio-SNe-compilation}
 \end{figure}

We compiled a sample of radio bright CCSNe from the literature in Figure ~\ref{Fig:radio-SNe-compilation} to compare the properties of \sn{} with other CCSNe, particularly SNe\,IIP. \sn{} occupies the lower luminosity part of the $\nu_{\rm p}$ $-$ $L_{\rm p}$ plot compared to other classes of SNe (Ib/c,IIn). However, it stands out as the brightest among SNe\,IIP. The dashed lines in the plot represent the mean velocities of the radio emitting shell if the peak flux is determined by SSA \citep{chevalier1998}. The inferred velocity of \sn{} is lowest among other SNe\,IIP and comparable to SNe\,IIn, which are known to have very dense CSM. If FFA is the dominant absorption process, the actual velocity will be larger than the ones deduced from this figure, which is indeed the case for \sn{} as shown in our study.  

\begin{figure} 
 	\centering
 	\includegraphics[width=0.45\textwidth]{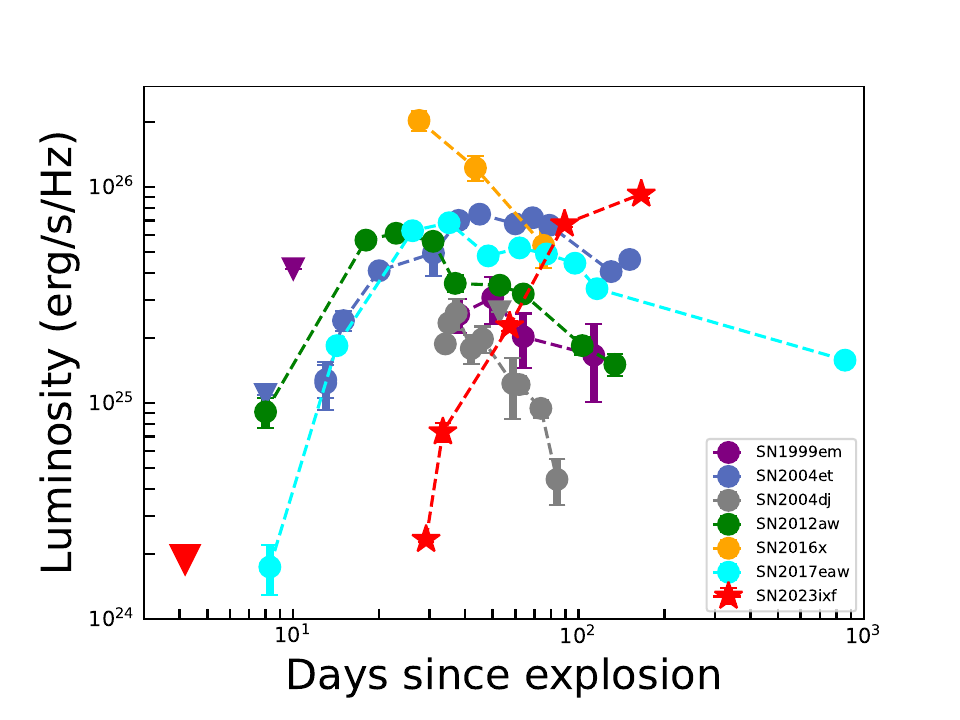} 
  \caption{The radio light curves (at $8-10$ GHz) of \sn{} (red star symbol) in context of other SNe\,IIP. While the radio luminosity of \sn{} is similar to that of other SNe\,IIP, the light curve rises significantly later than that of other SNe, indicating denser CSM. References: \cite{pooley2002}, \cite{chevalier2006}, \cite{nayana2018}, \cite{yadav2014}, \cite{carmona2022}. }
 \label{Fig:radio-lc-IIPs}
 \end{figure}

There are only six SNe\,IIP with detailed radio follow-up observations; SN\,1999em \citep{pooley2002}, SN\,2004et \citep{chevalier2006}, SN\,2004dj \citep{nayana2018}, SN\,2012aw \citep{yadav2014}, SN\,2016X \citep{carmona2022}, and SN\,2017eaw \citep{Szalai17eaw}. In Figure \ref{Fig:radio-lc-IIPs}, we present the $8-10$ GHz light curves of these SNe from the literature and compare them with that of \sn{}. The rise time of \sn{} is the slowest among the sample, reaching the peak of the light curve at $\delta t \approx 165$\,d. This indicates a relatively dense CSM for \sn{} compared to other SNe\,IIP. The mass-loss rates of the literature sample range from  $\dot M= 2.8 \times 10^{-6}\, \rm M_{\odot}\,yr^{-1}$ to $\dot M= 1.7 \times 10^{-5}\, \rm M_{\odot}\,yr^{-1}$, whereas the mass-loss rate of \sn{} is $\dot M \approx 10^{-4}\, \rm M_{\odot}\,yr^{-1}$ for an adopted wind velocity of $v_{\rm w}=25\,\rm km\,s^{-1}$ (value motivated by the high-resolution spectroscopy study of Dickinson et al., in prep.). This suggests that the progenitor of \sn{} was experiencing the most extreme mass-loss among the SNe\,IIP progenitors in this sample.
 
\begin{figure} 
 	\centering
 	\includegraphics[width=0.48\textwidth]{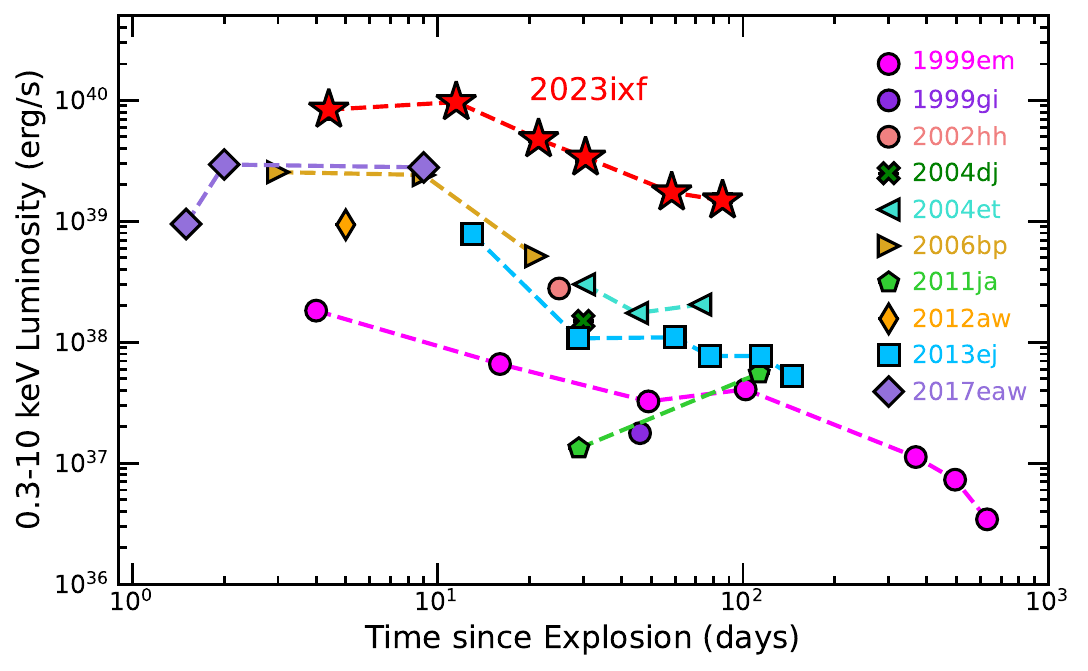}
         \caption{Unabsorbed 0.3-10 keV soft X-ray luminosity of the sample of detected type-IIP SNe. \sn{}\, clearly stands out as the most X-ray luminous type-IIP SN detected to date.  References: \cite{Schlegel94W,pooley2002,Schlegel99gi,Pooley02hh,Pooley04dj,Misra07,Immler07,Chakraborti11ja,Chakraborti13ej,Immler12aw,Szalai17eaw}. }
 \label{Fig:Xraycomparison}
 \end{figure}
We compiled the X-ray luminosity of all type-IIP SNe detected so far and find that \sn{} is the brightest among them (see Fig \ref{Fig:Xraycomparison}). Note that we do not plot SN2016X because it is unclear if the X-ray source faded or not \citep{Bose2019}, and the physical association of the X-ray emission with the SN is thus questionable. \sn{}\, stands out as the most X-ray luminous type-IIP SN found to date, and fills the part of the parameter space where of X-ray luminous IIP that were missing so far \citep{Dwarkadas14}.   

\sn{}\, has extreme radio and X-ray properties compared to the sample of known objects. Taken together, the late-time rising radio emission and the significantly luminous X-ray emission of \sn{} imply that the \emph{SN shock encountered a medium with properties that are not common among Type IIP SNe.} 
%---------------------------------------------------------------------------------
%-----------------------------------------------------------------------
\subsection{CSM density profile} \label{SubSec:DiscDensity}
\label{SubSec:csm-density-profile}
\begin{figure*} 
 	\centering
 	\includegraphics[width=0.9\textwidth]{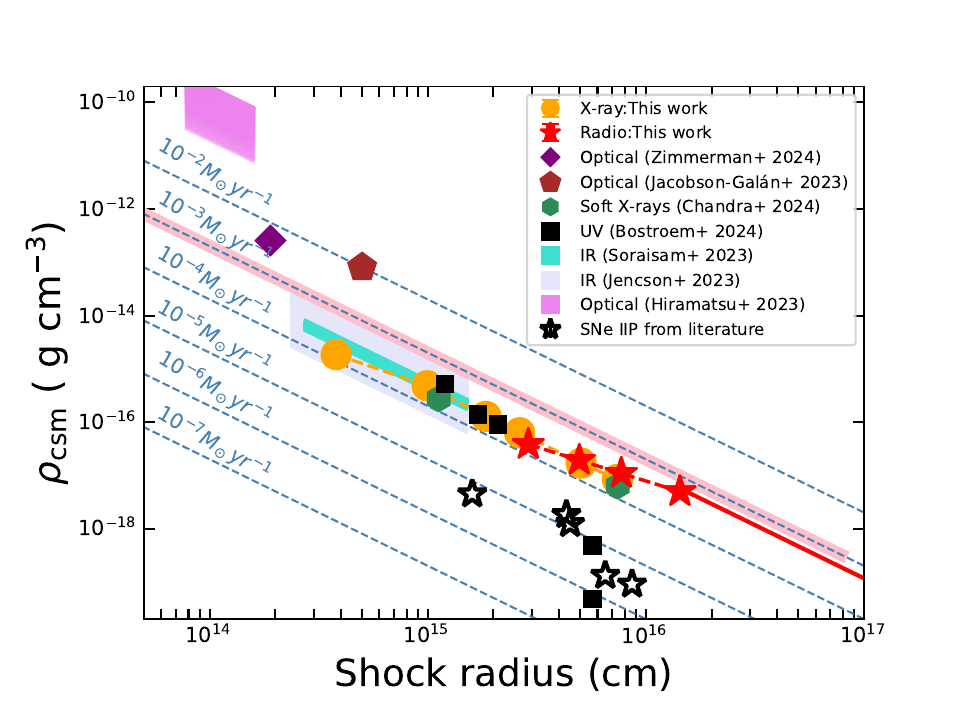} 
\caption{ Various estimates of CSM densities of \sn{}. The orange circles denote the densities derived from the X-ray emission measure (\ref{SubSec:densityXrays}), and red star symbols mark the densities from radio modeling (\ref{SubSec:multi-epoch radio modeling}) for $\epsilon_{\rm e}=10^{-4}$ and $T_{\rm e} =$ 5 $\times$ 10$^{5}$ K. The solid red line is the wind density profile corresponding to the best-fit optical depth in the radio model at $t \approx$ 165\,d. We also plot density estimates and limits from \citealt{Zimmerman23}, \citealt{JacobsonGalan23}, \citealt{Chandra23}, \citealt{Bostroem2024}, \citealt{Soraisam23}, \citealt{Jencson23}, and \citealt{Hiramatsu23}. The comparison of CSM densities derived from optical analysis with X-ray modeling implies a complex medium (asymmetry and/or clumps) at inner radii ($R<10^{15}$ cm). The mass-loss rates of other radio bright SNe\,IIP are marked as black open star symbols: SN\,1999em \citep{pooley2002}, SN\,2004dj \citep{nayana2018}, SN\,2004et \citep{chevalier2006}, SN\,2012aw \citep{yadav2014}, and SN\,2016X \citep{carmona2022}. These mass-loss rates are calculated assuming $v_{\rm w}=25\, \rm km/s$ and $v_{\rm sh}=10,000\, \rm km/s$. As a reference, typical mass-loss rates of RSGs are in the range $10^{-6}-10^{-5}\,\rm{M_{\odot}yr^{-1}}$ with $v_w=10-50\,\rm{km\,s^{-1}}$ \citep{Smith14}, with small fraction observed during the short-lived super-wind phase, which is characterized by mass-loss rates that are comparable to those of \sn{}\, $\dot M\approx 10^{-4}-10^{-3}\,\rm{M_{\odot}yr^{-1}}$. The pink line represents the average mass-loss rate of Galactic RSG VY Canis Majoris ($\dot M\approx 10^{-3}\,\rm{M_{\odot}yr^{-1}}$) for the last 10$^{3}$ years inferred from the HST images of the dust scattering nebula around the star \citep{Smith14}.}
 \label{Fig:csm-density-profile}
 \end{figure*} 
The main features of the CSM density profile of \sn{} derived from our combined X-ray and radio observations are the following:
(1) The CSM densities at shock radii $\approx$ 3.8 $\times$ 10$^{14}$  to 1.4 $\times$ 10$^{16}$ cm correspond to $\dot M\approx 10^{-4}\,\rm{M_{\odot}yr^{-1}}$ for a stellar wind velocity of $v_{\rm w} = 25\, \rm km\,s^{-1}$, which is at least an order of magnitude higher than typical RSG mass-loss rates \citep{Smith14,Beasor20}. 
(2) The intrinsic X-ray luminosity remains roughly constant for $\delta t <10$\,d and decays as $L_{\rm x} \propto t^{-1}$ at later epochs, indicating strong interaction with a dense CSM at $r<10^{15}$ cm. The $\rm NH_{\rm int}$ declines steeply as $\rm NH_{\rm int}(t) \propto t^{-2.3}$ at early epochs ($\delta t<10$\,d), and then follows a temporal evolution $\rm NH_{\rm int}(t) \propto t^{-1}$, consistent with a wind like CSM. However, the temporal evolution of $\rm NH_{\rm int}(t)$ inferred from the corresponding EM at $\delta t<10$\,d is significantly discrepant from the observed one and suggests that the neutral column fraction decreases from 70\% to 15\% between $\delta t \approx$ 4 to 11\,d. This may indicate asymmetries and/or clumpiness in the CSM at $r <10^{15}$ cm.
(3) The density profile derived from X-ray emission measure is wind-like at $R \approx (1-7.4) \times 10^{15}$ cm and the one from radio SED modeling is flatter ($\rho_{\rm CSM}(r) \propto r^{-1.27}$) at $R \approx (2.9-14) \times 10^{15}$ cm. If we extrapolate the density profile derived from radio modeling to early X-ray epochs, there is at least an order of magnitude difference in $\rho_{\rm CSM}(r)$ at $\delta t \approx 4$\,d and $\sim$ a factor of 5 difference at $\delta t \approx 11$\,d (see Figure~\ref{Fig:densities-diff-T}).

Various estimates of CSM densities of \sn{} at different distances from the explosion center are reported in the literature \citep{Berger23,JacobsonGalan23,Chandra23,Bostroem2024,grefenstette2023,Berger23,Neustadt2024,Soraisam23,Zimmerman23}. We compile all these density estimates and place them along with the densities derived from our study in Figure \ref{Fig:csm-density-profile} to build a comprehensive picture of the environment of \sn{}. 
\subsubsection{Progenitor in superwind phase of evolution}
\label{Subsec:Discussion-density-superwindphase}
The typical mass-loss rates of RSGs from empirical studies are $\dot{M} \approx 10^{-6}-10^{-5}\, \rm {M_{\odot}yr^{-1}}$ \citep{deJager1988,Nieuwenhuijzen1990,VanLoon2005,Beasor20,Smith14} where the high mass-loss rates \citep{VanLoon2005} are valid for dust enshrouded RSGs or maser emitters \citep{Goldman2017} that are not representative of early phase of RSG evolution. Observations of massive clusters indicate that RSGs with higher mass-loss rates are concentrated towards the end of the RSG phase, where pulsations and turbulence are important, driving them to blue-ward evolutionary tracks. Some of the most luminous RSGs tend to have higher mass-loss rates, which causes obscuration due to dust. An example of this is the Galactic object VY Canis Majoris, for which the HST images show dust scattering nebula whose density drops sharply at $\approx$ 8000 AU ($1.2 \times 10^{17}\, \rm cm$). The average mass-loss rate of this sytem is $\dot M\approx 10^{-3}\,\rm{M_{\odot}yr^{-1}}$ for last 10$^{3}$ years. High mass loss rates of $\dot M\approx (3-6) \times 10^{-4}\,\rm{M_{\odot}yr^{-1}}$ are reported for RSG IRC+10420 \citep{Knapp1985_IRC10420,Humphreys1997_IRC10420} and $\dot M\approx 10^{-4} - 10^{-3}\,\rm{M_{\odot}yr^{-1}}$ for NML Cygni \citep{Nagayama2008_NMLCyg,Lucas1992_NMLCyg}. The mass-loss rates derived for \sn{} from X-ray and radio observations are approximately an order of magnitude larger than that of other SNe\,IIP (see Fig \ref{Fig:csm-density-profile}) at comparable shock radii. This suggests that the progenitor is likely a very luminous RSG with high mass-loss rates, indicating a `superwind' phase of evolution with vigorous pulsations and turbulence. 

Pre-explosion archival mid-IR data of the site of \sn{} shows significant variability in the IR light curves with a periodicity of $\approx 1000$ d, indicating pulsational instability \citep{Soraisam23,Jencson23}. The initial mass of the progenitor candidate is estimated to be $\approx 17-20 M_{\odot}$ by comparing the luminosity with stellar evolution models, placing it as one of the most massive progenitors of SNe\,IIP to date. The progenitor mass-loss rates during 3$-$19 years before explosion from SED modeling is $\dot M\approx 3 \times 10^{-5} - 3 \times 10^{-4}\,\rm{M_{\odot}yr^{-1}}$ \citep{Jencson23} and from an empirical period–luminosity–based mass-loss prescription is $\dot M\approx (2-4) \times 10^{-4}\,\rm{M_{\odot}yr^{-1}}$ \citep{Soraisam23,Goldman2017}. The mass-loss rates derived from our X-ray and radio analyses correspond to $5-178$ years prior to the explosion and are in line with the values estimated from the pre-explosion variability studies.

Shock interactions with the CSM with a mass loss rate of $\dot{M}\sim({10}^{-4}-{10}^{-3})~M_\odot~{\rm yr}^{-1}$ also provide unique opportunities to investigate particle acceleration. 
High-frequency radio data, especially provided by NOEMA, can be explained by secondary leptons produced by CR protons via inelastic $pp$ interactions. Theoretically, not only electrons but also protons should be accelerated, and SNe such as SN 2023ixf can be regarded as the most promising targets for multimessenger observations, especially for nearby events~\citep{Murase2018}.

\subsubsection{Dense confined CSM}
\label{SubSec:Discussion-dense-confined-CSM}
\sn{} exhibits a dense, confined CSM with a mass-loss rate of $\dot M\approx 10^{-2}\,\rm{M_{\odot}yr^{-1}}$, which drops sharply to $\dot M\approx 10^{-4}\,\rm{M_{\odot}yr^{-1}}$ at $R \approx 10^{15}\, \rm cm$ \citep{JacobsonGalan23,Zimmerman23,grefenstette2023}. The intrinsic X-ray luminosity stays roughly constant up to $\delta t \approx 10$\,d and decays as $L_{\rm x} \propto t^{-1.1}$, indicating strong interaction with dense CSM at $r < 10^{15}$ cm. While these mass-loss rates are significantly higher than those observed in Galactic RSGs, there is no evidence of IR-bright pre-SN outbursts between 3 and 11 years ($R \approx 2.4-8.7 \times 10^{14}\, \rm cm$) before the explosion \citep{Jencson23}. IR observations up to $\approx 10$\,d before the explosion show no significant changes in colors or fluxes, ruling out the increased activity of progenitor star in the last 3 years as the cause of the dense confined CSM \citep{Jencson23}. Pre-explosion optical imaging studies also provide stringent constraints in this regard, showing no significant activity up to $\approx 400$\,d ($R \approx 8.64 \times 10^{13}\, \rm cm$) before the explosion \citep{Neustadt2024}. Pre-explosion eruptive activity was not detected in the Pan-STARRS light curves spanning 5000\,d prior explosion to a limiting magnitude of $-$7 \citep{Ransome2024}, which is below the brightness of other pre-SN outbursts in the literature. Thus, it is unlikely that the dense CSM around \sn{} is created due to an enhanced pre-SN eruption of the progenitor.

Shock waves launched near the photosphere of an RSG can support a dense region or chromosphere with similar densities that extend up to a few stellar radii. These chromospheric mass models can account for the observational signatures of dense confined CSM, such as flash-ionized emission lines at early times, early light curve peaks, and extended shock breakout emission, as observed in the case of \sn{} \citep{Fuller2024}. Therefore, the confined CSM in \sn{} is more likely due to an extended chromosphere rather than an eruptive pre-explosion outburst. In the chromospheric model, the shocks are generated by turbulent convection that varies stochastically, possibly resulting in a spatially asymmetric density structure. 

Non-thermal emission can also be used as a probe of dense confined CSM, because shock acceleration may operate after radiation breaks out from the dense CSM. Radio emission is unlikely to escape while the shock is embedded deep in the dense CSM, and higher-energy particles such as neutrinos and gamma-rays provide more promising signatures, as extensively studied for SN 2023ixf~\citep{Kheirandish2023,Guetta2023,Sarmah2024,Murase2024,Waxman2024,Kimura2024}.  

\subsubsection{Asymmetric and complex CSM}
\label{SubSec:Discussion-density-complex-csm}
There are several pieces of evidence suggesting asymmetric CSM in the immediate vicinity ($R < 10^{15}\, \rm cm$) of \sn{} from early continuum polarization studies \citep{Vasylyev23} and optical spectral analysis \citep{smith2023}. We tie down these observational signatures from the literature and the results from our analysis to investigate the complexity of the CSM up to distances $R \approx 1.4 \times 10^{16}$ cm from the explosion center. 

The CSM densities estimated from optical spectral modeling \citep{JacobsonGalan23} at $R \approx (0.5-1.5) \times 10^{15}$ cm are approximately two orders of magnitude higher than the densities derived from our X-ray spectral analysis (see Figure \ref{Fig:csm-density-profile}), indicating the co-existence of different densities at the same distance from the progenitor. The significantly steep evolution of $\rm NH_{\rm int}$ ($\rm NH_{\rm int} \propto t^{-2.3}$) at early times ($\delta t < 10$\,d) in contrast to that inferred from the EM also points to arguments on asymmetries and/or clumpiness at $r < 10^{15}$ cm. There is evidence for dense equatorial winds from RSGs, and numerical simulations of an SN interaction with such winds show that slight variation in angular density gradient will result in asymmetry in interaction shell greater than expected from a radial stellar wind \citep{Blondin1996}. Direct observations of RSGs also provide compelling evidence for asymmetric CSM environments$-$VY CMa \citep{Smith09VYCMa}, IRC+10420 \citep{Tiffany2010_IRC10420}, and NML Cygni \citep{Schuster2009_NMLCyg}. High-resolution CO spectra of the environment of RSG VY CMa indicate highly asymmetric and clumpy CSM created due to episodic mass-loss, possibly connected to pulsational or magnetohydrodynamic activity in the outer layers of the star \citep{Smith09VYCMa,Humphreys2021-VYCMa}. The resulting CSM could be asymmetric at distances up to $R \approx 10^{16}$ cm. 

The CSM densities of \sn{} derived from X-ray emission measure indicate a density profile $\rho_{\rm CSM}(r) \propto r^{-2}$ at $R \approx (1-7.4) \times 10^{15}\, \rm cm$. Radio spectral modeling implies a density profile $\rho_{\rm CSM}(r) \propto r^{-1.27}$ at $R \approx (2.9-14) \times 10^{15}\, \rm cm$. Densities derived from X-ray and radio modeling in the overlapping region $R \approx (2.9-7.4) \times 10^{15}\, \rm cm$ are consistent with each other for $T_{\rm e}=5 \times 10^{5}$ K (see Fig \ref{Fig:csm-density-profile}). However, if we extrapolate the density profile derived from radio modeling to early X-ray epoch, there is at least an order of magnitude difference at $t \approx$ 4\,d and $\sim$ a factor of 5 difference at $t \approx$ 11\,d. This could imply that the radio and X-ray emission are coming from two different regions of the CSM. However, the early radio non-detections are consistent with the possible radio emission that could arise from a synchrotron plasma that produces the observed X-ray emission. Thus, there is no strong evidence supporting the co-existence of two density profiles at $R \approx (0.4 - 7.4) \times 10^{15} \, \rm cm$. The plausible scenario could be the CSM density profile transitions from $\rho_{\rm CSM}(r) \propto r^{-2}$ to $r^{-1.27}$ at $R =(2.9-7.4)\times 10^{15}\, \rm cm$. An alternate explanation is the presence of clumpy CSM extending up to the radius probed by the radio observations. The density profile derived from radio modeling assumes a smooth CSM distribution. However, if the CSM is clumpy and has a significant structure such that the filling factor varies with the shock radius (more to less clumpy with radii), the observed radio SEDs can be represented in a wind-like CSM profile. 

To summarize, the CSM of \sn{} is highly complex, shaped by the extreme mass-loss (effective $\dot{M} \approx 10^{-4} \rm M_{\odot}\,yr^{-1}$) of the progenitor up to around 178 years before the explosion, leading to asymmetric and clumpy density distribution extending from immediate environment ($R < 10^{15}$ cm) to distances $R \approx 10^{16}$ cm from the explosion site. Various possible time-dependent mass-loss mechanisms that could create such a density profile include nuclear instabilities, super Eddington winds, and Gravity waves \citep{Owocki2004,Owocki2017,Quataert2012,Quataert2016,Fuller2017,Wu-Fuller2021,Arnett2011,SmithArnett14}. 

%%%%%%%%%%%%%%%%%%%%%%%%%%%%%%%%%%%%%%%%%%%%%%%%%%%%%%%%%%%%%%%%%%%
\section{Conclusions} 
\label{Sec:conclusions}
We carried out extensive X-ray and radio follow-up observations of \sn{} over $\delta t \approx 4-165$\,d, covering energy bands 0.3--79 keV in the X-rays and frequency bands 0.7--25 GHz in the radio. We model the spectral and temporal evolution of these observations to investigate the properties of the SN shock and environment. The main findings from our study can be summarized as follows.

\begin{itemize}
\item The X-ray emission is absorbed thermal bremsstrahlung radiation dominated by the forward shock, where the level of absorption decreases with time. The intrinsic X-ray luminosity remains constant up to $\delta t \approx 10$\,d and decays as $L_{\rm x} \propto t^{-1.1}$ at later epochs.
\item The CSM density profile inferred from X-ray emission measure is wind-like at $R \approx (1-7.4) \times 10^{15}$ cm and deviates from a wind profile at $R < 10^{15}$ cm as also mirrored by the steep temporal decay of $\rm NH_{\rm int} \propto t^{-2.3}$ at $\delta t < 10$\,d. This can be reconciled by the presence of clumpy CSM in the line of sight at these radii ($R < 10^{15}$ cm).
\item Radio emission is non-thermal synchrotron radiation suppressed by FFA due to CSM of density profile $\rho_{\rm CSM}(r)\propto r^{-1.27}$ at $R \approx (3-14) \times 10^{15}$ cm. The difference in density profiles derived from X-ray and radio analyses could be attributed to a clumpy CSM with a variable filling factor. While various cooling processes could be important depending on the values of shock microphysical parameters ($\epsilon_{\rm e}$ and $\epsilon_{\rm B}$), the dominant FFA determines the CSM density profile and remains agnostic about the intrinsic shape of the SSA spectra. While late-time radio emission may be explained by synchrotron emission from primary electron acceleration, early-time radio fluxes at 84~GHz measured by NOEMA are not consistent with the extrapolated radio SEDs, especially with $p>2$. We show that this may be an indication of secondary leptons produced by CR protons via $pp$ interactions. To model the radio emission accurately, it is essential to systematically account for both primary and secondary components as well as various cooling losses in the electron energy distribution. 
\item The mass-loss rate inferred from X-ray and radio modeling is $\dot{M} \approx 10^{-4} \rm M_{\odot}\,yr^{-1}$ for a stellar wind velocity of $v_{\rm w} = 25 \rm km\,s^{-1}$ during $\approx 3-178$ years preceding the core-collapse. This rate is about an order of magnitude higher than those derived for other radio/X-ray bright SNe\,IIP in the literature and significantly exceeds the typical mass-loss rates of RSGs. This places the progenitor of \sn{} among those with the most extreme mass-loss rates. 
\end{itemize}
Our extensive multiwavelength observations and modeling of \sn{} paint a picture of extremely complex CSM with asymmetries and clumps from the immediate environment to $\approx 10^{16}$ cm from the explosion center. While various time-dependent mass-loss mechanisms can create such an intricate environment, our study demonstrates the need for detailed high-cadence follow-up observations using optical, X-ray, and radio facilities to fully map such a complex CSM structure. 
\section{Acknowledgments}.
\begin{acknowledgments}
%NOEMA
This work is based on observations carried out under project number d23ab with the IRAM NOEMA Interferometer [30m telescope]. IRAM is supported by INSU/CNRS (France), MPG (Germany) and IGN (Spain).
%NRAO
The National Radio Astronomy Observatory is a facility of the National Science Foundation operated under cooperative agreement by Associated Universities, Inc. 
%GMRT
We thank the staff of the GMRT that made these observations possible. The GMRT is run by the National Centre for Radio Astrophysics of the Tata Institute of Fundamental Research.
%NuSTAR
This research has made use of the NuSTAR Data Analysis Software (NuSTARDAS) jointly developed by the ASI Space Science Data Center (SSDC, Italy) and the California Institute of Technology (Caltech, USA). 
%XMM
Partially based on observations obtained with XMM-Newton, an ESA science mission
with instruments and contributions directly funded by
ESA Member States and NASA.
%Chandra
This research has made use of data obtained from the Chandra Data Archive provided by the Chandra X-ray Center (CXC).
%Leicester XRT +Italy XRT
This research has made use of the XRT Data Analysis Software (XRTDAS) developed under the responsibility
of the ASI Science Data Center (ASDC), Italy.
This work made use of data supplied by the UK Swift Science Data Centre at the University of Leicester.
%HEASARC
This research has made use of data and software provided by the High Energy Astrophysics Science Archive Research Center (HEASARC), which is a service of the Astrophysics Science Division at NASA/GSFC.
\end{acknowledgments}
%ASPEN
This work was performed in part at the Aspen Center for Physics, which is supported by National Science Foundation grant PHY-2210452.
% Theory network
The collaboration of DT and RM was facilitated by interactions that were funded by the Gordon and Betty Moore Foundation through Grant GBMF5076.
%Raf
R.M. acknowledges support by the National Science
Foundation under award No. AST-2221789 and AST-2224255, and by NASA under grant 80NSSC22K1587.
The TReX team at UC Berkeley is partially funded by the
Heising-Simons Foundation under grant 2021-3248 (PI: Margutti).
%Kohta
The work of K.M. is partially supported by the NSF Grants No. AST-2108466, No. AST-2108467, and No. AST-2308021, and KAKENHI No. 20H05852.
%Daichi
D. T. acknowledges support by the Sherman Fairchild Postdoctoral Fellowship at the California Institute of Technology.
%Poonam
P.C. acknowledges support from NASA through Chandra award No. DD3-24141X issued by the Chandra X-ray Center.
% Andrea Reguitti
AR acknowledges financial support from the GRAWITA Large Program Grant (PI P. D’Avanzo) and the PRIN-INAF 2022 ``Shedding light on the nature of gap transients: from the observations to the models".
%Sergio
SC thanks Norbert Schartel for granting the first {\it XMM-Newton} 
observation. SC acknowledges support from ASI.
%Marcote
BM acknowledges financial support from the State Agency for Research of the Spanish Ministry of Science and Innovation, and FEDER, UE, under grant PID2022-136828NB-C41/MICIU/AEI/10.13039/501100011033, and through the Unit of Excellence Mar\'ia de Maeztu 2020--2023 award to the Institute of Cosmos Sciences (CEX2019- 000918-M). The European Research Council (ERC) under the European Union’s Horizon 2020 research and innovation programme (`EuroFlash’: Grant agreement No.\ 101098079).
\vspace{5mm}
\facilities{HST(STIS), Swift(XRT and UVOT), AAVSO, CTIO:1.3m,
CTIO:1.5m, CXO}
\software{astropy \citep{2013A&A...558A..33A,2018AJ....156..123A},  
          %Cloudy \citep{2013RMxAA..49..137F}, 
          %Source Extractor \citep{1996A&AS..117..393B}
          }
\appendix
Tables \ref{Tab:radio-NOEMA}, \ref{tab:radio-vla}, and \ref{Tab:radio-gmrt} show radio observation logs and flux measurements of \sn{} with NOEMA, VLA, and GMRT, respectively. Table \ref{Tab:Xraylog} shows the details of X-ray observations of SN 2023ixf with XMM-Newton, CXO, and NuSTAR.

\section{Radio Data Table}\label{AppendixRadio}
\startlongtable
\begin{deluxetable*}{ccccccccc}
\tablecaption{NOEMA observations of SN\,2023ixf}
\tablehead{
\colhead{Start Date} & \colhead{Project ID}  & \colhead{Centroid MJD} & \colhead{Phase$^{\rm{a}}$} & \colhead{Frequency}   & \colhead{Flux Density$^{\rm{b}}$} \\
 (dd/mm/yy) &  &  & (d) & (GHz) & ($\mu$Jy) &
 }
\startdata
10/06/2023 & d23ab  &  60105.83 &  23.09  & 84 &  212$\pm$55  \\ 
26/06/2023 & d23ab  &  60121.04 &  38.30  & 84 &  359$\pm$62  \\ 
19/07/2023 & d23ab  &  60144.87 &  62.13  & 84 &  206$\pm$64  \\ 
09/06/2023 & d23ab  &  60104.11 &  21.37  & 144 & $<$200  \\ 
26/06/2023 & d23ab  &  60121.94 &  39.20  & 144 & $<$200  \\ 
18/07/2023 & d23ab  &  60143.93 &  61.19  & 144 & $<$365  \\ 
\enddata
\tablecomments{$^{\rm{a}}$ With respect to explosion. $^{\rm{b}}$ The uncertainties on flux measurements are 1$\sigma$, and upper limits are $5\sigma$.
\label{Tab:radio-NOEMA}}
\end{deluxetable*}

\startlongtable
\begin{deluxetable*}{ccccccccc}
\tablecaption{VLA observations of SN\,2023ixf}
\tablehead{
\colhead{Start Date} & \colhead{Project ID} & \colhead{VLA Config.} & \colhead{Centroid MJD} & \colhead{Phase$^{\rm{a}}$} & \colhead{Frequency} & \colhead{Bandwidth} & \colhead{Flux Density$^{\rm{b}}$} \\
 (dd/mm/yy) &  &  &  & (d) & (GHz) & (GHz) & ($\mu$Jy) &
 }
\startdata
23/05/2023 & SF151070 &  B  &  60087.01 &  4.18  &   10     & 2.0 &  $<$ 33 $\mu$Jy \\
17/06/2023 & SF151070 &  A  &  60112.04 & 29.21  &  10      & 2.0 & 41 $\pm$ 8 $\mu$Jy \\
21/06/2023 & SF151070 & A  &  60116.30 &  33.45 &   8.5	   & 0.5 &	51  $\pm$   12  \\ 
21/06/2023 & SF151070 & A  &  60116.30 &  33.45 &    9.5   & 0.5 &  109  $\pm$   11  \\
21/06/2023 & SF151070 & A  &  60116.30 &  33.45 &    10.5  & 0.5 &  129  $\pm$   15  \\
21/06/2023 & SF151070 & A  &  60116.30 &  33.45 &    11.5  & 0.5 &  134  $\pm$   16  \\
21/06/2023 & SF151070 & A  &  60116.30 &  33.45 &    12.5  & 0.5 &  154  $\pm$   14  \\
21/06/2023 & SF151070 & A  &  60116.30 &  33.45 &    13.5  & 0.5 &  198  $\pm$   17  \\
21/06/2023 & SF151070 & A  &  60116.30 &  33.45 &    14.5  & 0.5 &  211  $\pm$   18  \\
21/06/2023 & SF151070 & A  &  60116.30 &  33.45 &    15.5  & 0.5 &  229  $\pm$   19  \\
21/06/2023 & SF151070 & A  &  60116.30 &  33.45 &    16.5  & 0.5 &  267  $\pm$   21  \\
21/06/2023 & SF151070 & A  &  60116.30 &  33.45 &    17.5  & 0.5 &  304  $\pm$   23  \\
21/06/2023 & SF151070 & A  &  60116.30 &  33.45 &    18.5  & 0.5 &  316  $\pm$   23  \\
21/06/2023 & SF151070 & A  &  60116.30 &  33.45 &    19.5  & 0.5 &  330  $\pm$   23  \\
21/06/2023 & SF151070 & A  &  60116.30 &  33.45 &    20.5  & 0.5 &  293  $\pm$   24  \\
21/06/2023 & SF151070 & A  &  60116.30 &  33.45 &    21.5  & 0.5 &  290  $\pm$   29  \\
21/06/2023 & SF151070 & A  &  60116.30 &  33.45 &    22.5  & 0.5 &  333  $\pm$   31  \\
21/06/2023 & SF151070 & A  &  60116.30 &  33.45 &    23.5  & 0.5 &  310  $\pm$   29  \\
21/06/2023 & SF151070 & A  &  60116.30 &  33.45 &    24.5  & 0.5 &  334  $\pm$   26  \\
21/06/2023 & SF151070 & A  &  60116.30 &  33.45 &    25.5  & 0.5 &  287  $\pm$   29  \\
\hline
15/07/2023  & SF151070 & A & 60140.14 &   57.31  &     2.5  & 0.5 &   $<$45  \\ 
15/07/2023  & SF151070 & A & 60140.14 &   57.31  &     3.5  & 0.5 &   $<$45   \\
15/07/2023  & SF151070 & A & 60140.14 &   57.31  &     4.5  & 0.5 &   82  $\pm$ 14  \\
15/07/2023  & SF151070 & A & 60140.14 &   57.31  &     5.5  & 0.5 &   131 $\pm$ 20  \\
15/07/2023  & SF151070 & A & 60140.14 &   57.31  &     6.5  & 0.5 &   192 $\pm$ 21  \\
15/07/2023  & SF151070 & A & 60140.14 &   57.31  &     7.5  & 0.5 &   269 $\pm$ 17  \\
15/07/2023  & SF151070 & A & 60140.14 &   57.31  &     8.5  & 0.5 &   345 $\pm$ 12  \\
15/07/2023  & SF151070 & A & 60140.14 &   57.31  &     9.5  & 0.5 &   393 $\pm$ 21  \\
15/07/2023  & SF151070 & A & 60140.14 &   57.31  &     10.5 & 0.5 &   399 $\pm$ 28  \\
15/07/2023  & SF151070 & A & 60140.14 &   57.31  &     11.5 & 0.5 &   439 $\pm$ 28  \\
15/07/2023  & SF151070 & A & 60140.14 &   57.31  &     12.5 & 0.5 &   517 $\pm$ 31  \\
15/07/2023  & SF151070 & A & 60140.14 &   57.31  &     13.5 & 0.5 &   550 $\pm$ 32  \\
15/07/2023  & SF151070 & A & 60140.14 &   57.31  &     14.5 & 0.5 &   517 $\pm$ 30  \\
15/07/2023  & SF151070 & A & 60140.14 &   57.31  &     15.5 & 0.5 &   549 $\pm$ 35  \\
15/07/2023  & SF151070 & A & 60140.14 &   57.31  &     16.5 & 0.5 &   530 $\pm$ 30  \\
15/07/2023  & SF151070 & A & 60140.14 &   57.31  &     17.5 & 0.5 &   549 $\pm$ 37  \\
15/07/2023  & SF151070 & A & 60140.14 &   57.31  &     18.5 & 0.5 &   515 $\pm$ 28  \\
15/07/2023  & SF151070 & A & 60140.14 &   57.31  &     19.5 & 0.5 &   539 $\pm$ 37  \\
15/07/2023  & SF151070 & A & 60140.14 &   57.31  &     20.5 & 0.5 &   545 $\pm$ 40  \\
15/07/2023  & SF151070 & A & 60140.14 &   57.31  &     21.5 & 0.5 &   481 $\pm$ 50  \\
15/07/2023  & SF151070 & A & 60140.14 &   57.31  &     22.5 & 0.5 &   442 $\pm$ 37  \\
15/07/2023  & SF151070 & A & 60140.14 &   57.31  &     23.5 & 0.5 &   457 $\pm$ 43  \\
15/07/2023  & SF151070 & A & 60140.14 &   57.31  &     24.5 & 0.5 &   442 $\pm$ 35  \\
15/07/2023  & SF151070 & A & 60140.14 &   57.31  &     25.5 & 0.5 &   402 $\pm$ 31  \\
\hline
16/08/2023  &   SF151070 & A & 60172.13 &  89.30 &    2.5 & 0.5 &   $<$45   \\  
16/08/2023  &   SF151070 & A & 60172.13 &  89.30 &    3.5 & 0.5 &   125 $\pm$  16  \\
16/08/2023  &   SF151070 & A & 60172.13 &  89.30 &    4.5 & 0.5 &   338 $\pm$  25  \\
16/08/2023  &   SF151070 & A & 60172.13 &  89.30 &    5.5 & 0.5 &   645 $\pm$  39  \\    
16/08/2023  &   SF151070 & A & 60172.13 &  89.30 &    6.5 & 0.5 &   854 $\pm$  47  \\
16/08/2023  &   SF151070 & A & 60172.13 &  89.30 &    7.5 & 0.5 &  1023 $\pm$  55  \\
16/08/2023  &   SF151070 & A & 60172.13 &  89.30 &    8.5 & 0.5 &  1182 $\pm$  60  \\    
16/08/2023  &   SF151070 & A & 60172.13 &  89.30 &    9.5 & 0.5 &  1236 $\pm$  66  \\
16/08/2023  &   SF151070 & A & 60172.13 &  89.30 &   10.5 & 0.5 &  1183 $\pm$  68  \\
16/08/2023  &   SF151070 & A & 60172.13 &  89.30 &   11.5 & 0.5 &  1259 $\pm$  79  \\
16/08/2023  &   SF151070 & A & 60172.13 &  89.30 &   12.5 & 0.5 &  1337 $\pm$  71  \\
16/08/2023  &   SF151070 & A & 60172.13 &  89.30 &   13.5 & 0.5 &  1190 $\pm$  65  \\
16/08/2023  &   SF151070 & A & 60172.13 &  89.30 &   14.5 & 0.5 &  1251 $\pm$  68  \\
16/08/2023  &   SF151070 & A & 60172.13 &  89.30 &   15.5 & 0.5 &  1193 $\pm$  64  \\
16/08/2023  &   SF151070 & A & 60172.13 &  89.30 &   16.5 & 0.5 &  1111 $\pm$  61  \\
16/08/2023  &   SF151070 & A & 60172.13 &  89.30 &   17.5 & 0.5 &  1110 $\pm$  63  \\
16/08/2023  &   SF151070 & A & 60172.13 &  89.30 &   18.5 & 0.5 &  1068 $\pm$  65  \\
16/08/2023  &   SF151070 & A & 60172.13 &  89.30 &   19.5 & 0.5 &  1052 $\pm$  62  \\
16/08/2023  &   SF151070 & A & 60172.13 &  89.30 &   20.5 & 0.5 &  1041 $\pm$  61  \\
16/08/2023  &   SF151070 & A & 60172.13 &  89.30 &   21.5 & 0.5 &  1068 $\pm$  65  \\
16/08/2023  &   SF151070 & A & 60172.13 &  89.30 &   22.5 & 0.5 &  1023 $\pm$  77  \\
16/08/2023  &   SF151070 & A & 60172.13 &  89.30 &   23.5 & 0.5 &   876 $\pm$  60  \\
16/08/2023  &   SF151070 & A & 60172.13 &  89.30 &   24.5 & 0.5 &   784 $\pm$  56  \\
16/08/2023  &   SF151070 & A & 60172.13 &  89.30 &   25.5 & 0.5 &   695 $\pm$  59  \\
\hline
31/10/2023 & SF151070 & D & 60248.61 &   165.78  &   2.5  & 0.5 &   $<$ 4593  \\
31/10/2023 & SF151070 & D & 60248.61 &   165.78  &   3.5  & 0.5 &   $<$ 4393 \\
31/10/2023 & SF151070 & D & 60248.61 &   165.78  &   04.5  & 0.5 &   2032 $\pm$		170 \\
31/10/2023 & SF151070 & D & 60248.61 &   165.78  &   05.5  & 0.5 &   2155 $\pm$		146 \\
31/10/2023 & SF151070 & D & 60248.61 &   165.78  &   06.5  & 0.5 &   2111 $\pm$		157 \\
31/10/2023 & SF151070 & D & 60248.61 &   165.78  &   07.5  & 0.5 &   2060 $\pm$		137 \\
31/10/2023 & SF151070 & D & 60248.61 &   165.78  &   08.5  & 0.5 &   1923 $\pm$		124 \\
31/10/2023 & SF151070 & D & 60248.61 &   165.78  &   09.5  & 0.5 &   1826 $\pm$		119 \\
31/10/2023 & SF151070 & D & 60248.61 &   165.78  &   10.5  & 0.5 &   1627 $\pm$		109 \\
31/10/2023 & SF151070 & D & 60248.61 &   165.78  &   11.5  & 0.5 &   1605 $\pm$     107 \\
31/10/2023 & SF151070 & D & 60248.61 &   165.78  &   12.5  & 0.5 &   1599 $\pm$		122 \\
31/10/2023 & SF151070 & D & 60248.61 &   165.78  &   13.5  & 0.5 &   1549 $\pm$     102 \\
31/10/2023 & SF151070 & D & 60248.61 &   165.78  &   14.5  & 0.5 &   1402 $\pm$     116 \\
31/10/2023 & SF151070 & D & 60248.61 &   165.78  &   15.5  & 0.5 &   1367 $\pm$		 98 \\
31/10/2023 & SF151070 & D & 60248.61 &   165.78  &   16.5  & 0.5 &   1368 $\pm$		 98 \\
31/10/2023 & SF151070 & D & 60248.61 &   165.78  &   17.5  & 0.5 &   1279 $\pm$      75 \\
\enddata
\tablecomments{$^{\rm{a}}$ With respect to explosion. $^{\rm{b}}$ The uncertainties on flux measurements includes map rms values (1$\sigma$) and a 5\% systematic uncertainty on the flux density added in quadrature. The flux density upper limits are $3\sigma$.
\label{tab:radio-vla}}
\end{deluxetable*}

\startlongtable
\begin{deluxetable*}{ccccccccc}
\tablecaption{GMRT observations of SN\,2023ixf}
\tablehead{
\colhead{Start Date} & \colhead{Project ID}  & \colhead{Centroid MJD} & \colhead{Phase$^{\rm{a}}$} & \colhead{Frequency}  & \colhead{Bandwidth} & \colhead{Flux Density$^{\rm{b}}$} \\
 (dd/mm/yy) &  &  & (d) & (GHz) & (GHz) & ($\mu$Jy) &
 }
\startdata
22/05/2023 & 44$_{-}$095  &  60086.00 &  3.17 & 1.25 & 0.28 & $<$ 75  \\ 
22/05/2023 & 44$_{-}$095  &  60086.00 &  3.17  & 0.75 & 0.20 & $<$ 200  \\ 
28/05/2023 & 44$_{-}$095  &  60092.00 &  9.17  & 1.25 & 0.28 & $<$ 80  \\ 
31/05/2023 & 44$_{-}$095  &  60095.00 &  12.17  & 0.75 & 0.20 & $<$ 200  \\ 
02/07/2023 & 44$_{-}$095  &  60127.00 &  44.17  & 1.25 & 0.28 & $<$ 68  \\ 
29/07/2023 & 44$_{-}$095  &  60154.00 &  71.17  & 1.25 & 0.28 & $<$ 200  \\ 
29/07/2023 & 44$_{-}$095  &  60154.00 &  71.17  & 0.75 & 0.20 & $<$ 75  \\ 
07/09/2023 & 44$_{-}$095  &  60194.00 &  111.17  & 0.75 & 0.20 & $<$ 200  \\ 
08/09/2023 & 44$_{-}$095  &  60195.30 &  112.47  & 1.25 & 0.28 & $<$ 70 \\ 
31/10/2023 & 45$_{-}$091  &  60248.00 &  165.17  & 1.25 & 0.28 & 148 $\pm$ 31  \\ 
\enddata
\tablecomments{$^{\rm{a}}$ With respect to explosion. $^{\rm{b}}$ The uncertainties on flux measurements are 1$\sigma$, and upper limits are $3\sigma$.
\label{Tab:radio-gmrt}}
\end{deluxetable*}

%%%%%%%%%%%%%%%%%%%%%%%%%%%%%%%%%%%%%%%%%%%%%%%%%%%%%%%%%%%%%%%%%%%%%%%%%%%%%%%%%%%
\section{X-ray observations logs}\label{AppendixXray}

\startlongtable
\begin{deluxetable*}{lcccccc}
\tablecaption{X-ray observations of \sn{} with \xmm, \chandra, and \nustar}
\tablehead{
\colhead{Instrument} & \colhead{Start date} & \colhead{Mid time$^{\rm{a}}$} & \colhead{Obs ID} & \colhead{Exposure Time$^{\rm{b}}$} &PI \\
  & (dd/mm/yy) & (d) &  & (ks)  &  &
 }
\startdata
NuSTAR & 2023-05-22 &  4.38 & 90902520002 &  42.2/42.0   &B. Grefenstette\\
XMM/EPIC-pn & 2023-05-27 & 9.01 & 0931790101$^{\rm{c}}$ & 62.8  &S. Campana \\ 
XMM/EPIC-MOS1 & 2023-05-27 & 9.01 & 0931790101$^{\rm{c}}$ & 69.9  &S. Campana \\ 
XMM/EPIC-MOS2 & 2023-05-27 & 9.01 & 0931790101$^{\rm{c}}$ & 70.0 & S. Campana  \\ 
NuSTAR &  2023-05-29& 11.46 & 90902520004 & 42.0/41.6  & B. Grefenstette\\ 
CXO &  2023-05-31& 12.95 & 27862 & 20.1 &  P. Chandra\\
NuSTAR & 2023-06-08& 21.44& 90902520006 & 63.4/62.7 & B. Grefenstette\\ 
NuSTAR & 2023-06-17& 30.51& 80902505002& 56.1/55.6  &  R. Margutti \\
XMM/EPIC-pn & 2023-06-18 & 30.79 & 0921180101$^{\rm{d}}$ & 10.0 &  R. Margutti \\ 
XMM/EPIC-MOS1 & 2023-06-18 & 30.79 & 0921180101$^{\rm{d}}$ & 11.8 &  R. Margutti \\ 
XMM/EPIC-MOS2 & 2023-06-18 & 30.79 & 0921180101$^{\rm{d}}$ & 11.8 &  R. Margutti \\ 
XMM/EPIC-pn & 2023-07-15 & 58.16 & 0921180301$^{\rm{d}}$ & 13.4 &  R. Margutti \\
XMM/EPIC-MOS1 & 2023-07-15 & 58.16 & 0921180301$^{\rm{d}}$ & 18.4 &  R. Margutti \\
XMM/EPIC-MOS2 & 2023-07-15 & 58.16 & 0921180301$^{\rm{d}}$ & 18.7 &  R. Margutti \\
NuSTAR & 2023-07-15& 58.43 & 80902505004& 53.3/52.8 &  R. Margutti\\
CXO & 2023-08-11& 84.93 & 27862 & 10.5 &  P. Chandra\\
CXO & 2023-08-12&  85.67 & 28374 & 9.9 &  P. Chandra\\
\enddata
\tablecomments{$^{\rm{a}}$ With respect to explosion. $^{\rm{b}}$After excluding intervals of time with large background (e.g., due to proton flaring). For NuSTAR, we report the exposure times for module A and module B. $^{\rm{c}}$ Observations acquired with the thick filter. $^{\rm{d}}$ Observations acquired with the medium filter. 
\label{Tab:Xraylog}}
\end{deluxetable*}

\bibliography{sn2023ixf_v1}{}
\bibliographystyle{aasjournal}

\end{document}